\documentclass[a4paper,10pt]{article}
\usepackage[paperheight=250mm,paperwidth=176mm,textheight=49pc,textwidth=136mm]{geometry}

\usepackage{graphicx}
\usepackage{epstopdf, epsfig}
\usepackage{amsmath}
\usepackage{amssymb}
\usepackage{graphicx}
\usepackage{subcaption}
\usepackage{natbib}
\usepackage{ulem}
\usepackage{comment}

\usepackage{xcolor}
\definecolor{darkred}   {rgb}{0.5 ,0   ,0   }
\definecolor{green}     {rgb}{0   ,0.8 ,0   }
\definecolor{darkblue}  {rgb}{0   ,0   ,0.5 }
\definecolor{greenblue} {rgb}{0   ,1   ,0.5 }
\definecolor{greenblue2}{rgb}{0.1 ,0.9 ,0.6 }
\definecolor{darkgreen} {rgb}{0.1 ,0.4 ,0.1 }
\definecolor{purple2}   {rgb}{0.72,0.27,1   }
\definecolor{magenta2}  {rgb}{1   ,0   ,1   }
\definecolor{gold}      {rgb}{0.9 ,0.6 ,0.1 }
\definecolor{red}       {rgb}{1,0,0}

\definecolor{legend_lightred}{HTML}{E36047}
\definecolor{legend_red}{HTML}{D14133}
\definecolor{legend_darkred}{HTML}{BB5C5B}
\definecolor{legend_lightblue}{HTML}{8CBDDB}
\definecolor{legend_blue}{HTML}{4D85BC}
\definecolor{legend_darkblue}{HTML}{5880B3}
\definecolor{legend_lightgreen}{HTML}{64BC72}
\definecolor{legend_darkgreen}{HTML}{00682A}

\usepackage{tikz}
\usetikzlibrary{arrows.meta} % in preamble

\newcommand{\mylab}[3]{\raisebox{#2}[0mm][0mm]{\makebox[0mm][l]{\hspace*{#1}#3}}}
\newcommand{\myarrow}[3]{\raisebox{#2}[0mm][0mm]{\makebox[0mm][l]{\hspace*{#1}
      \begin{tikzpicture}[overlay]
        #3
      \end{tikzpicture} } } }

\usepackage{hyperref}
\hypersetup{
    colorlinks=true,
    linkcolor=blue,
    filecolor=magenta,      
    urlcolor=red,
    citecolor=blue,
    pdfauthor={J.M.Catalan}
}

\title{Aerodynamic performance and robustness of a nature-inspired concept for a micro-scale
wind turbine}

\author{J. M. Catal\'an$^{1,2}$, G. Arranz$^{3}$, M. Moriche$^{4}$, M. Guerrero-Hurtado$^{1}$, \\ M. García-Villalba$^{4}$ and O. Flores$^{1,*}$  \\[2ex]
$^{1}$ Department of Aerospace Engineering \\ Universidad Carlos III de Madrid, Spain \\[1ex] 
$^{2}$ Simulation Technology Department \\ ITP Aero, Alcobendas, Spain \\[1ex] 
$^{3}$ California Institute of Technology, United States \\[1ex] 
$^{4}$ Institute of Fluid Mechanics and Heat Transfer \\ TU Wien, Vienna,  Austria \\[1ex]
$^{*}$ Corresponding author: oflores@ing.uc3m.es \\ [1ex]
}

 \date{}

\begin{document}

\maketitle


\begin{abstract}

We present direct numerical simulations of a novel concept for a micro-scale wind turbine, inspired in the mechanics of the auto-rotation of winged seeds. 
In this nature-inspired concept the turbine blades have two degrees of freedom: the pitch and the elevation (or coning) angles. 
These allow the blade to vary its attitude with respect to the incoming velocity seen by the blade (i.e., the tip-speed ratio, $\lambda$). 
In order to validate this new concept, we perform numerical simulations of the coupled fluid-solid problem, solving together the Navier-Stokes equations for the fluid and the Newton equations for the rigid body (i.e., the blade). 
We characterize a preliminary nature-inspired single-blade rotor over a range of operational conditions (including both uniform and turbulent inflows), 
demonstrating the ability of the novel rotor to extract power at a very low Reynolds number (i.e., $\mathrm{Re}=240$ based on the blade's chord and the freestream velocity), significantly changing its attitude in response to different braking torques and tip-speed ratios. 
The rotor achieves a peak power coefficient of $C_{P,\max} = 0.026$ at $\lambda \approx 2.0$. 
This peak value is unchanged between uniform and turbulence-perturbed inflows, demonstrating the robustness of the nature-inspired design. 
However, performance remains lower than that of fixed-blade configurations, showing that while the concept is feasible and stable, optimization of blade planform and mass distribution is essential to improve efficiency.

\end{abstract}

%===============================================================================
% Introduction
%===============================================================================
\section{Introduction} \label{sec:intro}

%The expansion of wireless sensor networks in sectors such as industry, agriculture, and environmental monitoring has been driven by advancements in the Internet of Things (IoT) technology, the promise of 5G connectivity, and the broader digitalization of society and economies. 
%
%These networks facilitate real-time data collection and transmission \citep{kandris2020applications}, with autonomously powered sensors playing a crucial role in their widespread deployment.
%
%While the most obvious power source for these devices seems to be solar cells, over the past decade there has been significant exploration into the development of compact wind turbines capable of operating at extremely low wind speeds, of the order of a gentle breeze.
%
%These turbines present a potential solution for micro-scale power generation across different environments, where solar cells might be compromised: dirty or dusty environments, nighttime operation or any other environment without sunlight (i.e. caves, systems of ducts, etc).

The rapid growth of wireless sensor networks in industry, agriculture, and environmental monitoring is fueled by advances in the Internet of Things (IoT), the deployment of 5G connectivity, and the ongoing digitalization of society. 
These networks rely on autonomous, long-lasting power sources to enable reliable real-time data collection and transmission \citep{kandris2020applications}.
While solar cells are the most common solution, they are ineffective in many practical scenarios, such as in dusty or shaded environments, at night, or in enclosed spaces like ducts and caves. 
To address these limitations, recent research has explored alternative micro-scale energy harvesters, including compact wind turbines capable of operating at extremely low wind speeds, on the order of a gentle breeze (i.e., $\lesssim 3$ m/s).

%Many investigations have been undertaken to enhance the efficiency of micro-scale wind turbines, with diameters smaller than 10cm.  
%
%Some of these works have focused on their aerodynamic design , whereas others have considered the integration of the miro-rotor with suitable energy converters and control modules  
%
%While the majority of research on rotor aerodynamics has concentrated on steady-state performance, few studies have focused on the performance of these devices in unsteady environments \citep{el2019stability}.
 
Significant efforts have been devoted to improving the efficiency of micro-scale wind turbines with diameters below 10 cm. 
These efforts have addressed both the aerodynamics of the micro-scale rotors 
\citep{leung2010design,kishore2013design,zakaria2015experimental,mendonca2017,ikeda2018robust,bourhis2022innovative},
and their integration with energy converters and control modules 
\citep{howey2011design,perez2016cm,gasnier2019}.
However, most aerodynamic studies have focused on steady-state performance, while only a few have examined the behavior of these devices under unsteady environments \citep{el2019stability}.

%Despite ongoing efforts, the feasibility of these compact wind turbines is yet to be fully demonstrated.
%
%At present, the efficiency of very micro-scale rotors remains relatively low, with power coefficient ($C_P$) values typically ranging from 0.05 to 0.35 \citep{bourhis2022innovative}. 
%
%For comparison, small, medium and large-scale wind turbines typically exhibit $C_P$ values in the range of 0.35-0.50, approaching the ideal maximum value of $C_{P,\max}\simeq 0.59$, known as the Betz limit \citep{betz1926wind}.
%
%The primary reason for this disparity lies in the challenges associated with the aerodynamic design of the micro-scale rotors. 
%
%Typically, very micro-scale rotors with diameters of the order of centimeters operate at small tip-speed ratios, in the range $\lambda \sim 0$ - $2$, where the tip-speed ratio is defined as the ratio of tip velocity $\Omega R$ over the freestream velocity $U_{\infty}$. 
%
%In this low tip-speed ratio regime, the wake rotation can produce a $C_{P,\max}$ reduction of up to $30\%$ with respect to the Betz limit for $\lambda=1$ \citep{wood2015maximum}.
%
%Likewise, the aerodynamic efficiency can be significantly reduced by the presence of relatively thick boundary layers and laminar separation bubbles, due to the low-Reynolds number associated to the low relative wind speed of the blades. 
%
%Also, the low speed and micro-scale of these rotors make them more vulnerable to gusts and freestream turbulence, whose characteristic length and intensity are typically comparable to those of the rotor, resulting in very unsteady operational environments.
Despite considerable research, the feasibility of compact wind turbines remains uncertain. 
Current micro-scale rotors exhibit relatively low efficiencies, with power coefficients ($C_P$) in the range of 0.05--0.35 \citep{bourhis2022innovative}. 
Typical $C_P$ values of large scale turbines are 0.35--0.50, and the theoretical maximum given by the Betz limit is $C_{P,\max} \simeq 0.59$ \citep{betz1926wind}. 
The main reason for this disparity lies in the aerodynamic limitations at small scales. 
Centimeter-scale rotors usually operate at low tip-speed ratios ($\lambda \sim 0$--$2$), defined as the ratio of tip velocity to the free-stream velocity. 
When $\lambda = 1$, wake rotation can reduce $C_{P,\max}$ by up to 30\% relative to the Betz limit \citep{wood2015maximum}. 
Moreover, the aerodynamic efficiency of the blade is also reduced by the thick boundary layers and laminar separation bubbles caused by the small relative wind speed, resulting in very low Reynolds numbers. 
Finally, the reduced size and slow rotation rates make these devices highly susceptible to gusts and ambient turbulence, whose characteristic scales are often comparable to those of the rotor itself, leading to unsteady operating conditions.

In this work, we propose to leverage the knowledge provided by nature and propose to investigate the feasibility of a micro-scale rotor based on the auto-rotational flight of winged seeds like samaras. 
%
%To this aim, we test the aerodynamic performance of a rotor made of blades that mimic the shape and motion of a free-falling samara.
%
These seeds, consisting of a nut attached to a wing-like rigid membranous structure, exhibit high aerodynamic efficiency stemming from the tight coupling between its inertia and the aerodynamic forces \citep{lugt1983autorotation}, resulting in a passively regulated autorotation that is triggered at very small velocities.
Similar to a helicopter, this rotation generates a lift force that counteracts the weight of the seed, reducing the descent speed and allowing larger dissemination by lateral winds.
As a matter of fact, this parachuting effect is one of the main reasons why winged seeds have garnered significant interest from the scientific community over the past few decades \citep{norberg1973autorotation,green1980terminal,azuma1989flight,rosen1991vertical,seter1992stability,yasuda1997autorotation, arranz2018kinematics}. 

Samaras are particularly interesting for the design of micro-scale wind turbines for several reasons. 
First, their autorotation is highly robust, occurring across specimens with diverse shapes and sizes \citep{schaeffer2024maple}. 
This robustness suggests that the underlying aerodynamic mechanism can be effectively emulated in engineered devices, a concept supported by previous work on samara-inspired aerial vehicles \citep{fregene2010dynamics, obradovic2012multi, ulrich2010falling}.
Second, the autorotation is remarkably stable even in the presence of flow disturbances \citep{lee2017flight, varshney2011kinematics}, with studies indicating that this stability is closely linked to the formation of a leading-edge vortex (LEV) over the wing 
\citep{lentink2009leading,salcedo2013stereoscopic,lee2014mechanism,arranz2018numerical}. 
Third, the characteristic size and descent velocity of samaras are comparable to the dimensions and operating speeds targeted by micro-scale wind turbines, making them excellent candidates for nature-inspired design.

From a mechanical perspective, the autorotation arises from the interplay between aerodynamic forces and moments generated by the wing and the inertial forces associated with the rotation of the seed. 
Numerical simulations using simplified wing planforms and mass distributions have shown that the angular velocity and attitude of a samara (i.e., its pitching and coning angles) vary systematically with descent speed 
\citep{arranz2018kinematics}. 
This  suggests that a micro-scale rotor with freely hinged blades could passively adjust its attitude in response to changes in wind speed, potentially improving the operating range of the rotor and the maximum power coefficient.
The objective of the present study is to explore the feasibility of such nature-inspired rotor and evaluate its performance in uniform and turbulent free-stream conditions.

The rest of the manuscript is structured as follows: section \ref{sec:methodology} presents the methodology, focusing on the governing equations, the rotor model and the computational setup.
Section \ref{sec:results} presents first the simulation results for a single-blade rotor in a uniform freestream, discussing the rotor performance and the flow structures surrounding the blade. Then, the single-blade rotor is tested in a freestream with turbulence-like perturbations, and then compared with a fixed attitude rotor.  
Finally, section \ref{sec:conclusions} draws some conclusions, describing the advantages and limitations of our novel concept of bioinspired micro wind turbine.

%===============================================================================
%===============================================================================
\section{Methodology}\label{sec:methodology}
%===============================================================================
%===============================================================================

\subsection{Numerical Methods}\label{sec:num-methods}

We solve numerically the fluid-structure interaction problem between the rigid bioinspired rotor and a Newtonian fluid of constant density $\rho$ and kinematic viscosity $\nu$.
The evolution of the fluid velocity $\mathbf u$ and pressure $p$ is governed by the Navier-Stokes equations for an incompressible flow:
\begin{subequations}\label{eq:ns}
    \begin{align}
\nabla \cdot \mathbf u \, = & \, 0, \label{eqns:a} \\
\frac{\partial \mathbf u}{\partial t} + (\mathbf u \cdot \nabla) \mathbf u \, = & \, - \nabla p + \frac{1}{\mathrm{Re}}\nabla^2\mathbf u + \mathbf f, \label{eqns:b}
\end{align}
\end{subequations} 
which have been nondimensionalized with the freestream flow velocity $U_{\infty}$, the fluid density $\rho$ and the maximum chord of the rotor's blade, $c$. 
Hence, the Reynolds number is $\mathrm{Re}=U_{\infty}c/\nu$.
The volumetric force $\mathbf f$ in eq. \eqref{eqns:b} is the direct forcing of the Immersed Boundary Method (IBM) used to impose the no-slip and impermeability boundary conditions on the rotor's surface \citep{uhlmann2005}. 

The motion of the rotor is governed by the 
Newton-Euler equations, which can be reduced to \citep{featherstone2014rigid}: 
\begin{equation}\label{eq:mb}
    \mathrm{H}(\mathbf q) \ddot{\mathbf{q}} + \mathrm C(\mathbf q, \dot{\mathbf q}) = \mathbf \xi + \mathbf \xi_a ,
\end{equation}
where $\mathbf q$ is the vector with the degrees of freedom of the rotor, 
$\mathrm H(\mathbf q)$ is the generalized inertia matrix or joint space, 
$\mathrm C(\mathbf q, \dot{\mathbf q})$ is a vector with the generalized bias force (i.e, Coriolis, centrifugal and gravity forces), 
$\mathbf \xi$ is the vector of generalized forces at the joints 
and $\mathbf \xi_a$ represents the vector of generalized aerodynamic forces.  

Equations \eqref{eq:ns} and \eqref{eq:mb} are solved together using the algorithm described in \cite{arranz2022fluid}. 
The incompressible Navier-Stokes equations with the IBM are solved using a fractional-step method, employing a uniform cartesian grid and second order finite differences for the spatial discretization of the fluid variables. 
The terms in eq. \eqref{eq:mb} are computed using the recursive algorithms of \cite{felis2017rbdl}. 
The two system of equations are advanced in time using a 3-step semi-implicit Runge-Kutta scheme, that provides a weak coupling between eqs. \eqref{eq:ns} and \eqref{eq:mb}.  
The solver is implemented in Python using Numba and Nvidia CUDA libraries to run on a GPU card. 
More details regarding the algorithm and its implementation can be found in \citet{arranz2022fluid} and \citet{guerrero2024gpu}, respectively.  

%================================================
\subsection{Rotor model} \label{sec:rotor-model}
%================================================

The rotor is composed by a single blade ($N_b = 1$) that rotates azimuthally with respect to an axis parallel to the incoming flow.
The effects of other blades, structural elements (hub, tower, etc) and gravity are ignored, i.e. the blade is the only body interacting with the freestream flow.
Power extraction is modelled with the application of a braking torque, as explained at the end of this section. 

\begin{figure}[t]
\centering
($a$)
\includegraphics[width=0.46\columnwidth]{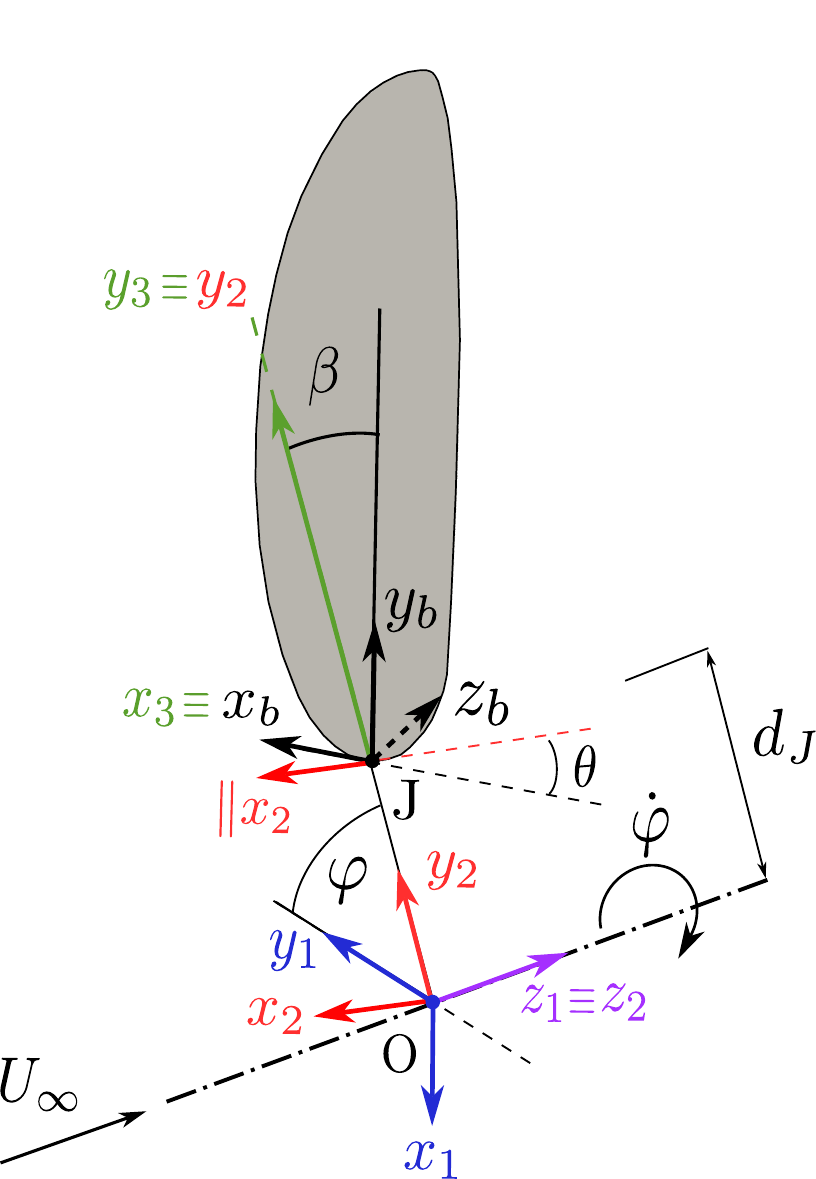}
\hspace{1mm} ($b$)
\includegraphics[width=0.40\columnwidth]{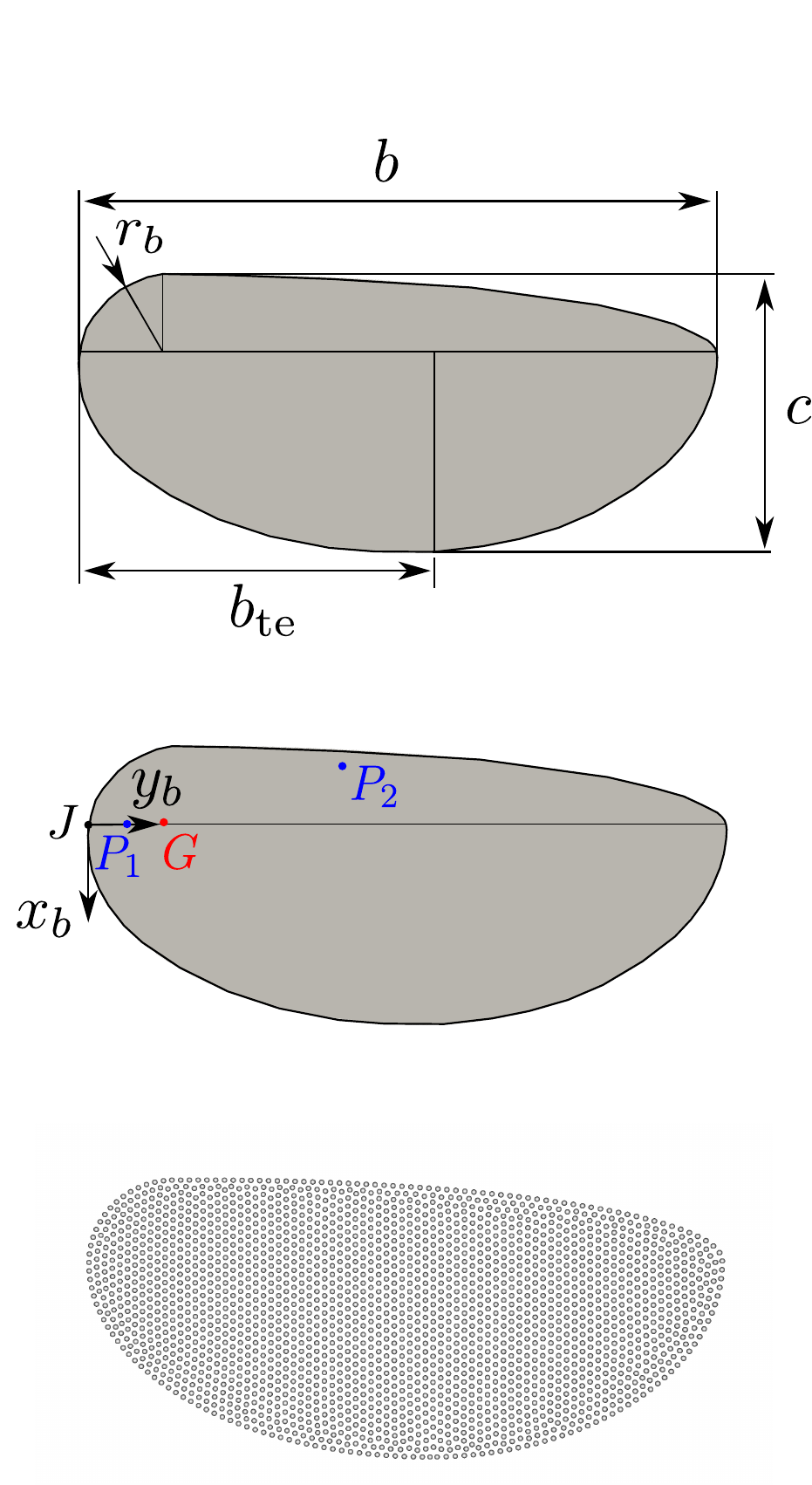}
\caption{
($a$) Sketch of the rotor model and reference frames (not at scale).  
$\Sigma_1$ in {\color{blue} blue}, 
$\Sigma_2$ in {\color{red} red}, 
$\Sigma_3$ in {\color{darkgreen} green}, 
and $\Sigma_b$ in black. 
($b$) Planform geometry of the blade, with explicit indication of dimensions (upper); body reference frame, 
center of mass 
and point masses locations (center); and blade discretization (lower). 
}
\label{fig:samaraWT-model}
\end{figure}

As discussed in the introduction, apart from the azimuthal rotation of the rotor ($\varphi$), the blade has two additional degrees of freedom: 
the pitch angle $\theta$ %around the radial direction ($y_2$), 
and the coning (or elevation) angle $\beta$. %around the chordwise axis $x_b$. 
These angles are shown in figure \ref{fig:samaraWT-model}a, which includes the four
reference frames used to define them. 
The inertial laboratory frame is $\Sigma_1$, center at $O$, with the incoming freestream velocity $U_\infty$ parallel to the $z_1$-axis. 
The azimuthal rotation of the rotor ($\varphi$) around the $z_1$-axis results in the second (non-inertial) reference frame, $\Sigma_2$.  
The plane of the rotor is then defined by either $(x_1,y_1)$ or $(x_2,y_2)$, and $z_1 \equiv z_2$ is the axis of rotation of the turbine. 
The blade is articulated at point $J$, located in the 
$y_2$-axis at a distance $d_J=0.3c$ from $O$. 
The locus of point $J$ is a circle in the rotor plane, and its position is determined by $\varphi$. 
We define two additional reference frames ($\Sigma_3, \Sigma_b$) centered at point $J$, which result from performing two consecutive rotations.
The reference frame $\Sigma_3$ is the result of a pitching rotation $\theta$ around the $y_2$ axis. 
Then, the body reference frame $\Sigma_b$ is the result of a coning rotation $\beta$ around the $x_3 \equiv x_b$-axis. 
The chordwise and spanwise directions are $x_b$ and $y_b$, respectively.  

The blade is designed following the idealized samara of \citet{arranz2018kinematics}. 
It is a flat surface contained in the $z_b=0$ plane, with constant thickness ($h=0.0035c$), no camber and no geometric twist. 
As shown in figure \ref{fig:samaraWT-model}b, the blade's shape is   defined by a quarter of a circle and three quarters of ellipse. 
The radius of the circle is $r_b=0.3c$, where $c$ is the blade's chord.  
The major and minor axis of the three ellipses can be defined in terms of $c$, $r_b$, the blade's span $b=2.2c$, and the spanwise distance to the trailing edge maximum $b_{\mathrm{te}}=0.54b=1.188c$.

The blade's freedom to rotate at $J$ (pitch and coning) is the key ingredient of the nature-inspired micro-rotor's design,  allowing the blade to adjust its attitude in response to changes in freestream velocity, rotation speed and/or braking torque. 
This response is controlled by the mass distribution on the blade, which again, is similar to the idealized samara of \cite{arranz2018kinematics}:
a uniform density on the flat plate $\rho_b$, and two point masses to model the mass of the nut (point $P_1$ in figure \ref{fig:samaraWT-model}b) and the mass of the leading edge nerve (point $P_2$). 
The density ratio between the flat plate's material and the fluid is $\rho_b/\rho=300$, which corresponds to the density ratio of a low density composite in air.
This yields a plate with a mass equal to $1.81\rho c^3$.
The point $P_1$ is located at $(0, 0.2c, 0)$ in $\Sigma_b$, with a mass of $20.36 \rho c^3$. 
It is the largest contributor to the total mass of the blade. 
The point $P_2$ is located at $(-0.2c, 0.9c, 0)$ in $\Sigma_b$, with a mass of $5\rho c^3$. 
The total mass of the blade (flat plate plus point masses) is then $27.17\rho c^3$, with the center of mass at $(-0.00255c, 0.38709c,0)$ in $\Sigma_b$ (see point G in figure \ref{fig:samaraWT-model}b). 
The tensor of inertia of the blade with respect to the center of mass is 
\begin{equation}
\bar{\bar{I}}_G =     
\begin{bmatrix}
3.4605 & 0.2720 & 0  \\
0.2720 & 0.3543 & 0  \\
0      & 0      & 3.8148
\end{bmatrix}
\rho c^5,
\end{equation}
computed in the body reference frame $\Sigma_b$ as well.

To extract power, a braking torque $T_b= -\rho U_\infty c^4 \, c_0 \dot{\varphi}$ is applied to the rotor axis, $z_1$. 
This torque is proportional to the rotor angular velocity ($\dot \varphi = d\varphi /dt$), with a dimensionless proportionality constant $c_0 > 0$. 
When the rotor achieves a steady state, the braking torque and the torque produced by the aerodynamic forces on $z_1$ must be in equilibrium, i.e. $T_b + M_{z,1} = 0$. 
Hence, we define the non-dimensional power coefficient as
\begin{equation}\label{eq:cp}
    C_P = \frac{2 \, \dot \varphi \, M_{z,1}}{\rho \, U_{\infty}^3 \, A_{\mathrm{disk}}},
\end{equation}
where $A_{\mathrm{disk}}=\pi(R_{\mathrm{max}}^2 - R_{\mathrm{min}}^2)$ is the maximum frontal area swept by the rotor, with $R_{\mathrm{max}}=2.5c$ and $R_{\mathrm{min}}=0.3c$.
Note that this area is different from the actual instantaneous frontal area, that would be dependent on the coning angle $\beta$. 
However, this simplifies the comparison between the different cases, normalizing $C_P$ with the maximum available power in the freestream.

%================================================
\subsection{Computational setup}\label{sec:computational-setup}
%================================================

\begin{figure}
\centering
\includegraphics[width=0.7\columnwidth]{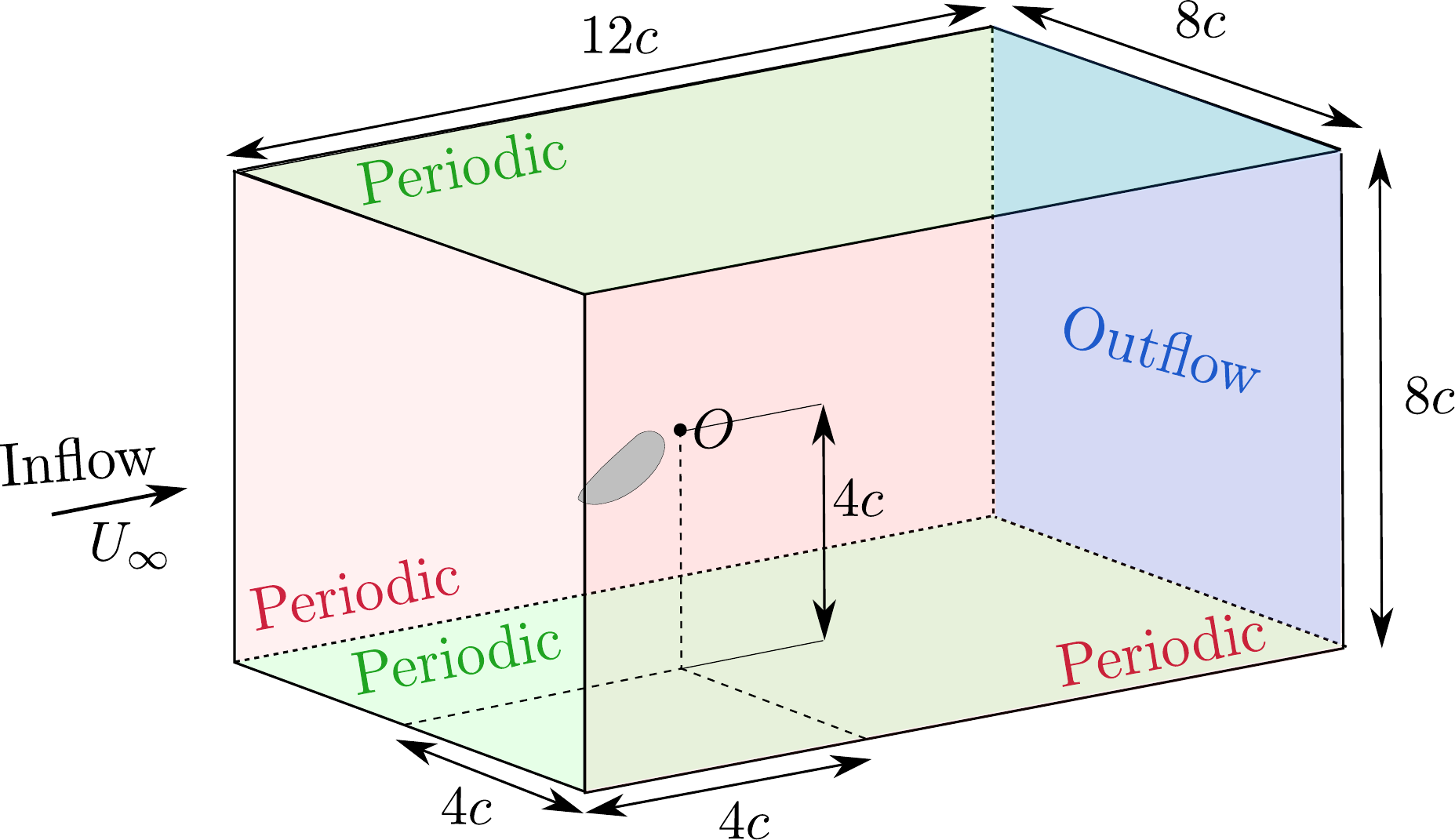}
\caption{Three-dimensional sketch of the computational domain and the boundary conditions.
\vspace*{-0.5cm}}
\label{fig:computational-domain}
\end{figure}

Given the similitude in geometry and configuration, the computational setup for the present micro-scale wind turbine is analogous to the one used in \citet{arranz2018kinematics} to simulate winged seeds in free fall. 
The micro-scale rotor is placed in the Cartesian computational domain shown in figure \ref{fig:computational-domain}.
We chose an inflow-outflow configuration, with a uniform inlet velocity $U_\infty$ imposed at $z_1 = -4c$, and an advective boundary condition applied at $z_1 = 8c$. The lateral boundary conditions are periodic, with a domain size $L_x = L_y = 8c$. 
The computational domain is discretized with a uniform Cartesian grid with $384\times384\times576$ points, which corresponds to a grid spacing of 48 grid points per chord, $\Delta_x = c/48$ (same as \citealp{arranz2018kinematics}). This grid spacing has been verified for the present configuration with a grid refinement study (see Appendix \ref{sec:grid-convergence}). 
The same grid spacing is used to discretize the blade, as shown in figure \ref{fig:samaraWT-model}b. Note that since the blade thickness ($h=0.0035c$) is much smaller than $\Delta_x$, we follow \cite{arranz2018kinematics} and model the blade as surface (i.e., a planar distribution of Lagrangian IBM points). 
The time step is set to $\Delta t \simeq 4\cdot10^{-4} c/U_\infty$, ensuring that the maximum CFL number is $\lesssim 0.1$.

For the simulations with turbulent perturbations in the freestream (see section \ref{sec:turbulence}), we use the synthetic turbulence generator of \citet{Schmidt2017source}. It consists of an additional volumetric force (to be added to the right hand side of eq. \ref{eqns:b}) with the following form, 
\begin{equation} \label{eq:fstig}
    \mathbf{f}_{\mathrm{st}}(x_1,y_1,z_1,t) = \frac{\mathbf{u}'(x_1,y_1,t)}{\Lambda_0/U_\infty} \exp\left(-\dfrac{\pi}{2} \dfrac{\left(z_1-z_{1,\mathrm{st}}\right)^2}{\Lambda_0^2} \right),
\end{equation}
where $\mathbf u'$ is the perturbation velocity, $z_{1,\mathrm{st}}$ is the position of the injection plane and $\Lambda_0$ is the integral length-scale of the velocity perturbations. 
The velocity fluctuation vector $\mathbf u'$ is a function of the cross flow coordinates ($x_1,y_1$) and time, and it is characterized by its (turbulent) intensity, $TI = \langle \| \mathbf u' \|^2 \rangle^{1/2}/U_\infty$.  
In this study, we generate $\mathbf u'$ using the digital filter approach of \citet{Klein2003digital} and \citet{Kempf2005efficient}. 
As discussed in \citet{catalan2024generation}, this procedure introduces isotropic velocity perturbations in the flow with a narrow energy spectra peaking at wavelengths around $\Lambda_0$ (see also \citet{catalan2024low}).
These velocity fluctuations are advected downstream, transitioning  into a flow analogous to the classic grid-induced turbulence in a relatively short development length (i.e., about 5-10$\Lambda_0$).

%\clearpage
%===============================================================================
% Results
%===============================================================================
\section{Results} \label{sec:results}

We perform simulations of the single-blade nature-inspired rotor described in section \ref{sec:rotor-model}, using the numerical methods described in section \ref{sec:num-methods} at a fixed Reynolds number $\mathrm{Re}=U_\infty c / \nu = 240$. 
This Reynolds number is the same used in the idealized winged seeds of \citet{arranz2018kinematics}, and corresponds to a  micro-scale wind turbine with $c=0.5$cm (i.e., $b=1.1$cm) in a freestream with velocity velocity of $U_\infty = 0.7$m/s  and kinematic viscosity $\nu = 1.46 \cdot 10^{-5}$ m$^2$s$^{-1}$ (i.e., air in standard sea level conditions).

\subsection{Nature-inspired micro-rotor in a uniform freestream}
\label{sec:unif-freestream}

\begin{figure}[t]
\includegraphics[width = 0.47\textwidth]{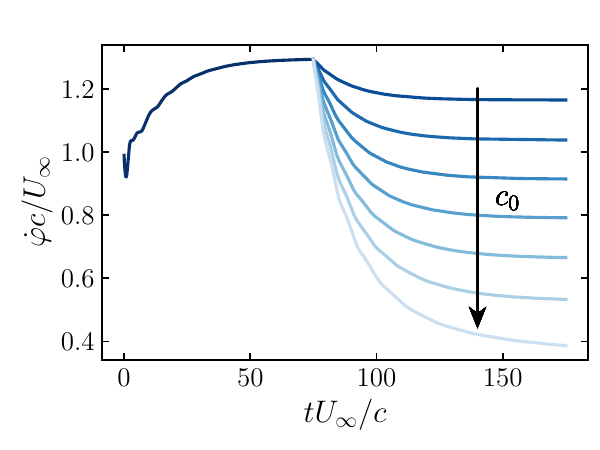}
\includegraphics[width = 0.47\textwidth]{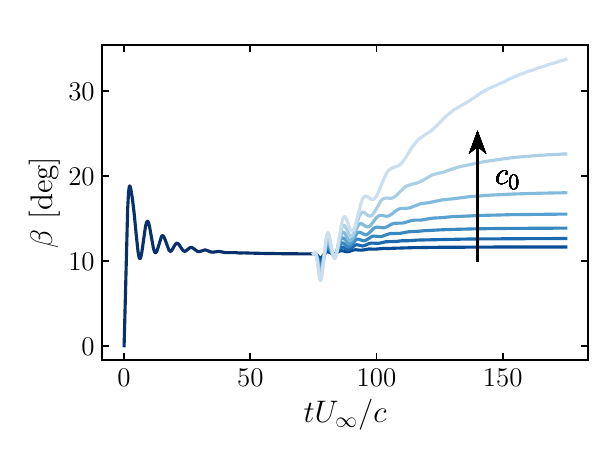}
\mylab{-0.94\textwidth}{0.33\textwidth}{$(a)$}
\mylab{-0.47\textwidth}{0.33\textwidth}{$(b)$}

\includegraphics[width = 0.47\textwidth]{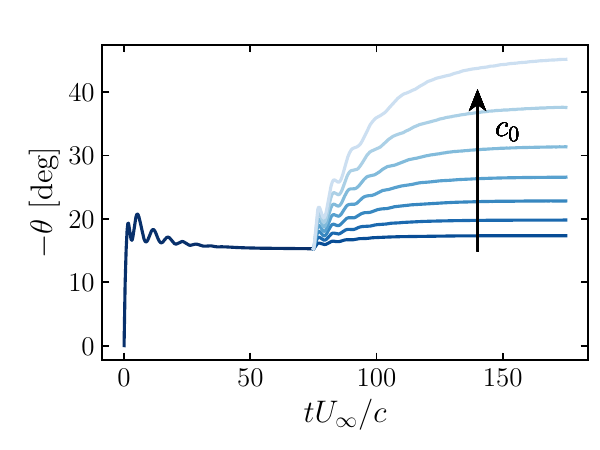}
\includegraphics[width = 0.47\textwidth]{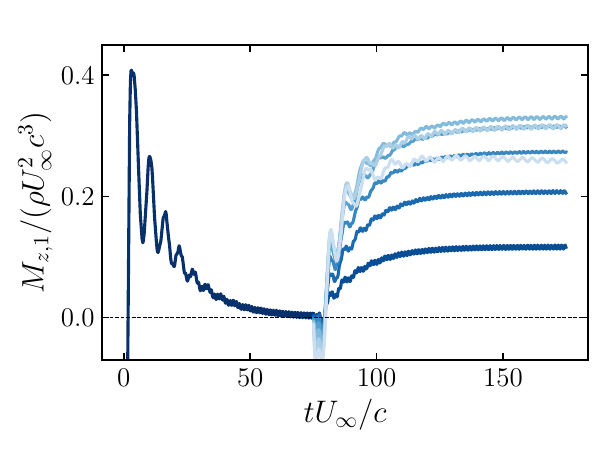}
\mylab{-0.94\textwidth}{0.33\textwidth}{$(c)$}
\mylab{-0.47\textwidth}{0.33\textwidth}{$(d)$}

\caption{Temporal evolution of the blade's kinematics and aerodynamic torque with a uniform freestream. 
($a$) Angular velocity, $\dot \varphi$; 
($b$) coning angle, $\beta$; 
($c$) pitching angle, $\theta$; and 
($d$) aerodynamic torque at the axis, $M_{z,1}$. 
The colors correspond to different values of the proportionality constant of the braking torque $c_0$, increasing in the direction of the arrows. 
\label{fig:1blade_time}
}
\end{figure} 

We consider first the nominal case of the nature-inspired, single-blade rotor extracting energy from a uniform freestream. 
Figure \ref{fig:1blade_time} shows the temporal evolution of the 3 degrees of freedom of the blade and the aerodynamic torque applied at the rotor's axis, $M_{z,1}$. The time span $0 < tU_\infty/c < 75$ corresponds to the initialization of the rotor without braking torque, with $c_0=0$, for about 13 revolutions of the rotor.  
The initial conditions at $t=0$ are $(\varphi, \beta, \theta) = (0, 0, 0) $ and $(\dot \varphi, \dot \beta, \dot \theta) = (U_\infty/c, 0, 0)$ for the blade's attitude and angular velocity, respectively. 
Figure \ref{fig:1blade_time}a shows that without braking torque the angular velocity of the blade increases from the initial condition to achieve a stable auto-rotation condition, in which the aerodynamic torque on the axis (and hence the power extracted by the turbine) is zero (see figure \ref{fig:1blade_time}d).  
The steady auto-rotation values for the coning (figure \ref{fig:1blade_time}b) and pitching (figure \ref{fig:1blade_time}c) angles are $\beta = 10.8^\circ$ and $\theta = -15.3^\circ$, 
with an angular velocity $\dot\varphi = 1.28U_\infty/c$ that yields a tip-speed ratio $\lambda = \dot\varphi  R_{\max} / U_\infty = 3.2$, where $R_{\max} = d_J + b = 2.5c$ is the distance from the axis to the blade tip when $\beta = \theta = 0$.
These values are very similar to those reported by \cite{arranz2018kinematics} for the free-falling winged-seed used to design our blade, namely $\dot\varphi = 1.58 U_\infty/c, \beta = 10^\circ$ and $\theta = -13^\circ$. 
This suggests that the restrictions to the blade attitude introduced by fixing $d_J\ne 0$ and restricting the 2 degrees of freedom of the hinge at $J$ do not disturb significantly the dynamics of the winged seed in auto-rotation. 

\begin{figure}[t]
\includegraphics[width = 0.47\textwidth]{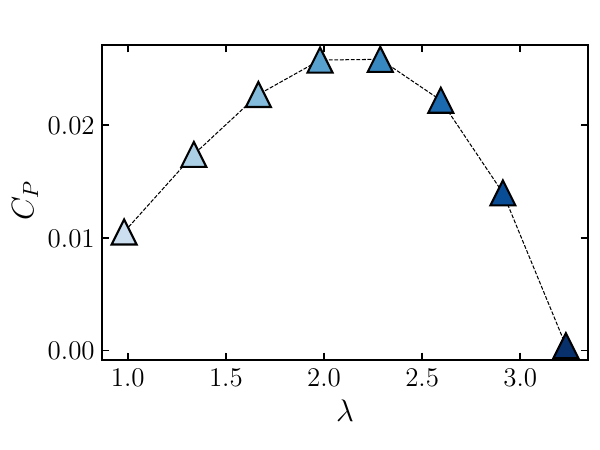}
\includegraphics[width = 0.47\textwidth]{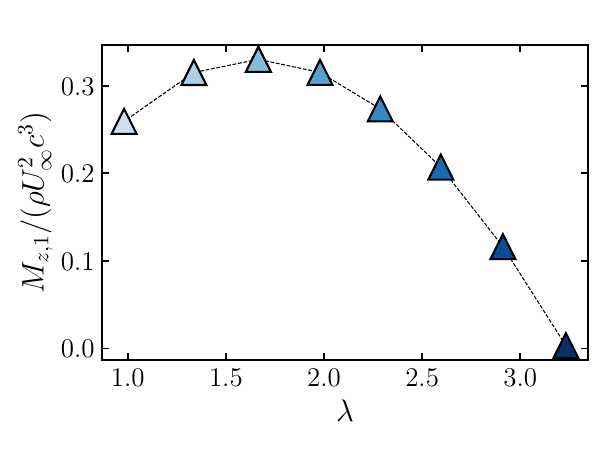}
\mylab{-0.94\textwidth}{0.33\textwidth}{$(a)$}
\mylab{-0.47\textwidth}{0.33\textwidth}{$(b)$}

\includegraphics[width = 0.47\textwidth]{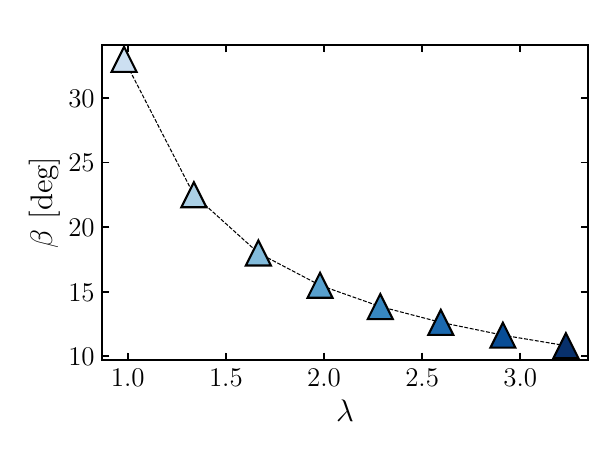}
\includegraphics[width = 0.47\textwidth]{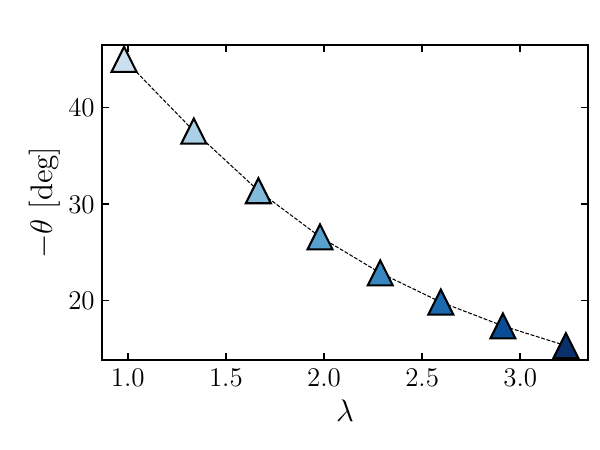}
\mylab{-0.94\textwidth}{0.33\textwidth}{$(c)$}
\mylab{-0.47\textwidth}{0.33\textwidth}{$(d)$}

\caption{Steady state performance and attitude of the rotor with a uniform freestream. 
($a$) Pressure coefficient,  
($b$) aerodynamic torque at the axis, 
($c$) coning angle and 
($d$) pitching angle, 
plotted as a function of the tip-speed ratio $\lambda = \dot\varphi R_{\max}/U_\infty$. 
The colors indicate the proportionality constant of the braking torque, as in figure \ref{fig:1blade_time}.
\label{fig:1blade_ST}
}
\end{figure}

The flow (velocity and pressure) and the blade's state (attitude and angular velocity) in auto-rotation (namely, at $t=75c/U_\infty$) are used as initial condition for the simulations of the blade with increasing values of the braking torque's proportionality constant,  $c_0 =0.1,\, 0.2,\, 0.3,\, 0.4,\, 0.5,\, 0.6$ and 0.7. 
These runs are shown in figure \ref{fig:1blade_time} for $t>75c/U_\infty$, using varying shades of blue. 
For each value of $c_0$, the application of the braking torque results in a gradual decrease of $\dot\varphi$ with time (see figure \ref{fig:1blade_time}a), reaching a quasi-steady state for $t \gtrsim 150 c/U_\infty$.
Only the case with the lowest angular speed (i.e., the largest braking torque proportionality constant, $c_0 = 0.7$) fails to reach a quasi-steady state, showing significant drifts in  $\dot\varphi, \beta, \theta$ and $M_{z,1}$ (see lighter blue curves in figure \ref{fig:1blade_time}).
Note that, as the blade decelerates (i.e.,  $\dot\varphi$ decreases with time), the blade's attitude changes, 
increasing $\beta$ (coning angle, figure \ref{fig:1blade_time}b) and  $-\theta$ (pitching angle, figure \ref{fig:1blade_time}c).  
As a result, the aerodynamic torque generated by the blade on the rotation axis increases (see figure \ref{fig:1blade_time}d), eventually balancing the braking torque applied at the axis that represents the energy generated by the rotor.  

It is interesting to note that, independently of the braking torque applied, the blade's attitude ($\beta$ and $\theta$) and the aerodynamic torque produced at the axis undergo high amplitude oscillations during a short transient,  $75 < tU_\infty/c \lesssim 100$. 
At the steady state (i.e., $t > 100 c/U_\infty$) the remaining oscillations have a relatively high angular frequency ($\approx 4 \dot\varphi$) and very low amplitude, specially for $\dot\varphi, \beta$ and $\theta$. 
The small steady-state oscillations can be marginally observed in the aerodynamic torque in figure \ref{fig:1blade_time}d, where the curve appears thicker than it is. 
Given their angular frequency, they are likely caused by the aerodynamic interference between {\it periodic} neighbors (recall that $x_1$- and $y_1$-directions are periodic). 
%\red{Juanma: queremos decir esto?} \olive{Oscar: yo diria que si ... al menos dejamos explicado en algun sitio este fenomenc}

The steady-state performance of the rotor is characterized using the time-averaged values of $\dot\varphi$, $\beta$, $\theta$ and $M_{z,1}$ during the last revolution of the blade. The time-averaged values of $\dot\varphi$ and $M_{z,1}$ allow computing the power coefficient $C_P$ using eq. \eqref{eq:cp}.  
These variables are plotted in figure \ref{fig:1blade_ST} as a function of the tip-speed ratio, $\lambda = \dot\varphi R_{\max}/U_\infty$.  
Figure \ref{fig:1blade_ST}a shows that the $C_p$ curve shows a shallow maximum, with a peak value of $C_P = 0.026$ when the tip-speed-ratio is $\lambda = 2.3$.  
The maximum torque at the axis is produced at a lower tip-speed-ratio, $\lambda = 1.7$ (see figure \ref{fig:1blade_ST}b). 
Note that the use of a braking torque proportional to $\lambda$ allows sampling the complete $C_P(\lambda)$  curve. 
Figure \ref{fig:1blade_ST}c and show how the blade transitions from the moderate values of $\beta$ and $\theta$ of auto-rotation (which correspond to an orientation close to perpendicular to the flow), to more extreme values at the lowest $\lambda$ (which corresponds to a blade that is becoming more parallel to the freestream).

Compared to the power coefficient of standard micro wind turbines with fixed angles, the performance of the present nature-inspired wind turbine is modest.
For example, 
\citet{gasnier2019} report $C_{P,\max} = 0.18 - 0.33$ and $\lambda_{\max} = 0.7 - 1.2$ for four-bladed micro wind turbines operating at Reynolds numbers in the range 1700-3400, 
while the  
three-bladed E62 micro wind turbine of \citet{mendonca2017} achieves $C_{P,\max} = 0.2$ at $\lambda_{\max} = 1$ with a larger $\mathrm{Re} \approx 10^4$.
% 
%Similarly, \ref{Howey2011} reports $C_{P,\max} \pprox 0.1$ and $\lambda_{\max} \\approx 1.1$ for a shrouded turbine with $N=12$ blade and a Reynolds number in the rante $600 < Re < 2000$. 
%
Normalizing performance by the number of blades (while neglecting aerodynamic interference and other losses) yields $C_{P, \max}^{N_b=1} = 0.045 - 0.08$, which is about two to three times higher than the $C_{P\max} = 0.026$ reported in figure \ref{fig:1blade_ST} for the present design.
However, this comparison requires some caution: the Reynolds number of our simulations ($\mathrm{Re}=240$) is significantly lower -- by a factor of two to three -- than that of the fixed-angle micro wind turbines, and our nature-inspired design has not yet been optimized. 
A one-to-one comparison between the nature-inspired rotor and fixed-angle rotors will be presented in section $\ref{sec:bio_vs_fix}$.

\begin{figure}[t]

\includegraphics[width=0.242\columnwidth, trim={100, 0, 100, 0},clip]{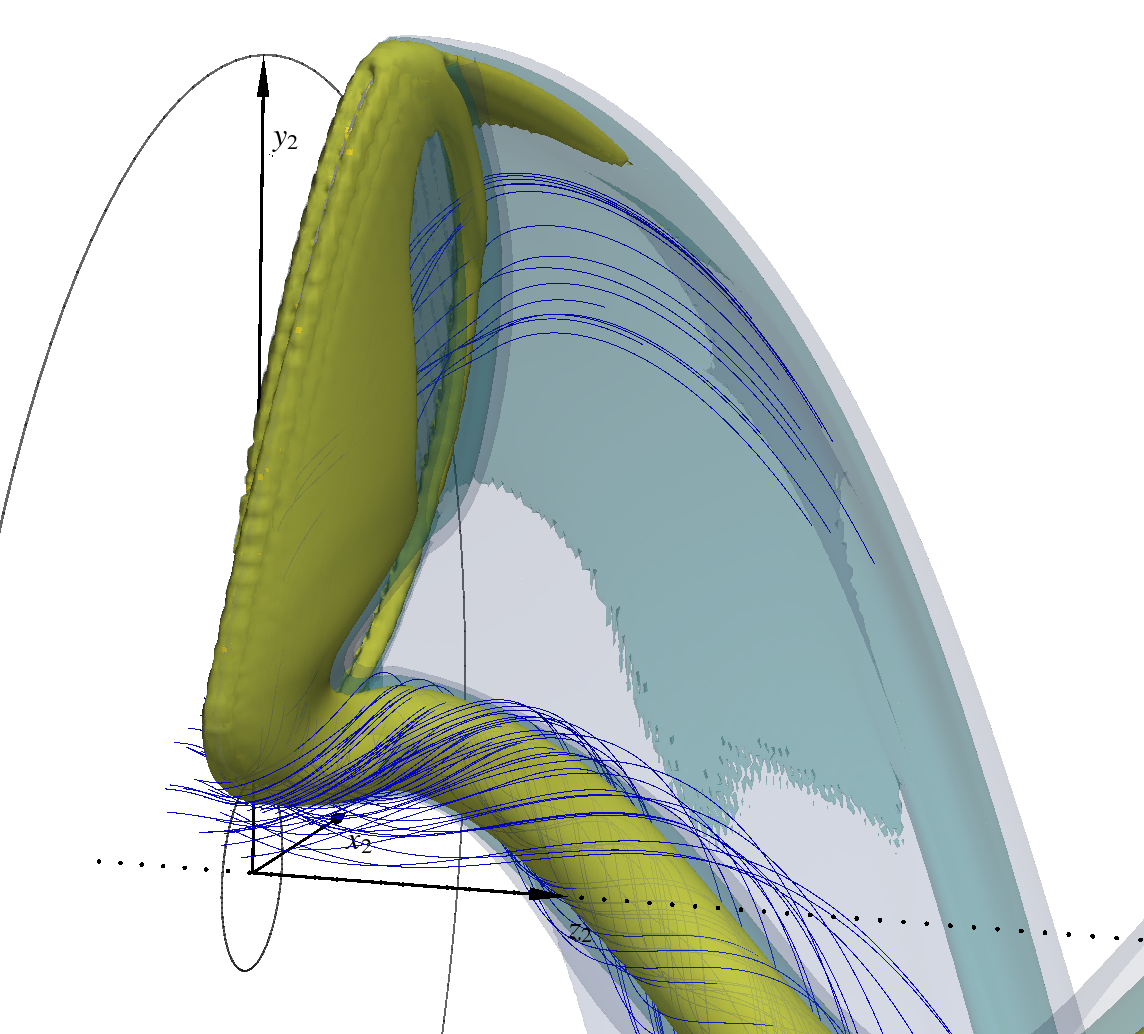}
\includegraphics[width=0.242\columnwidth, trim={100, 0, 100, 0},clip]{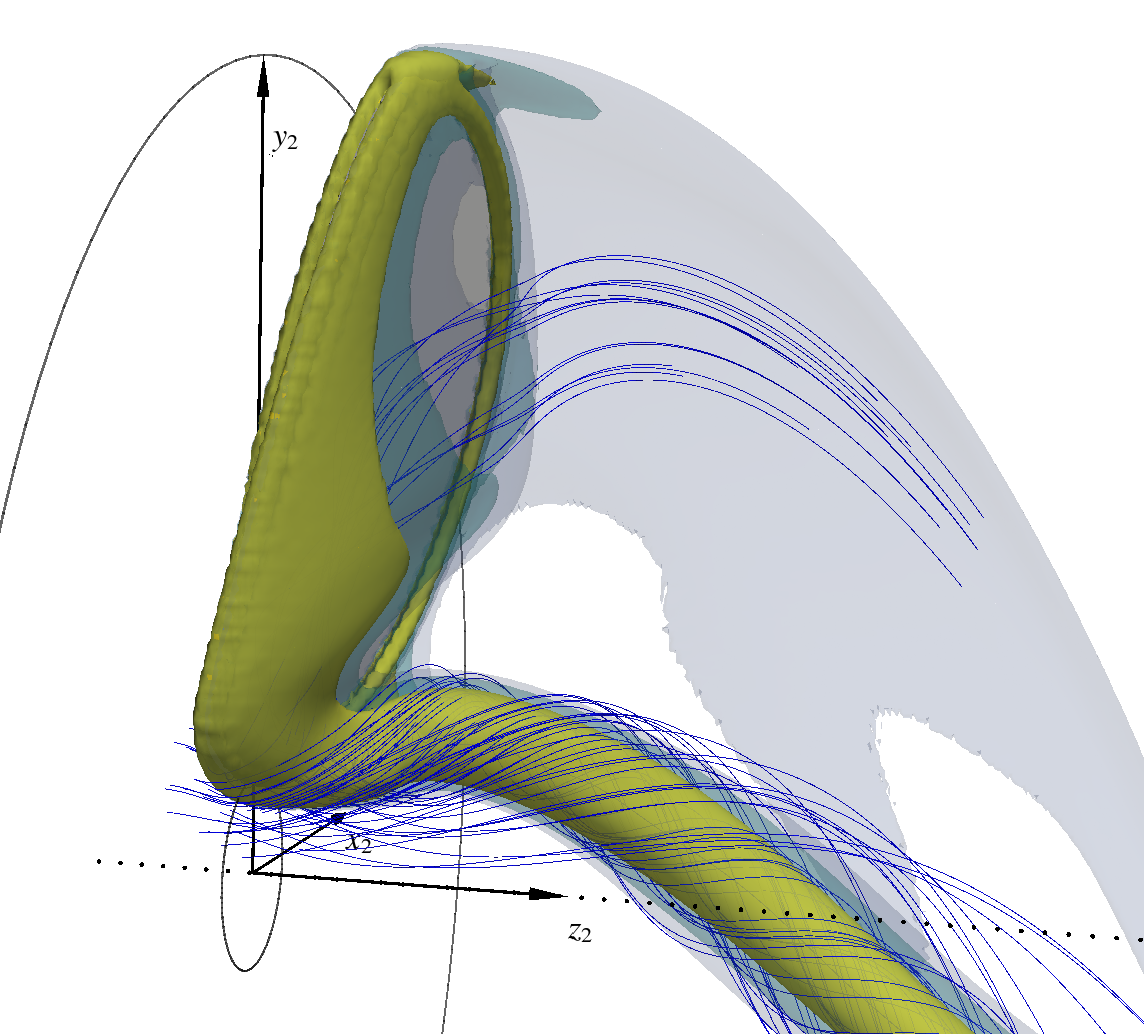}l
\includegraphics[width=0.242\columnwidth, trim={100, 0, 100, 0},clip]{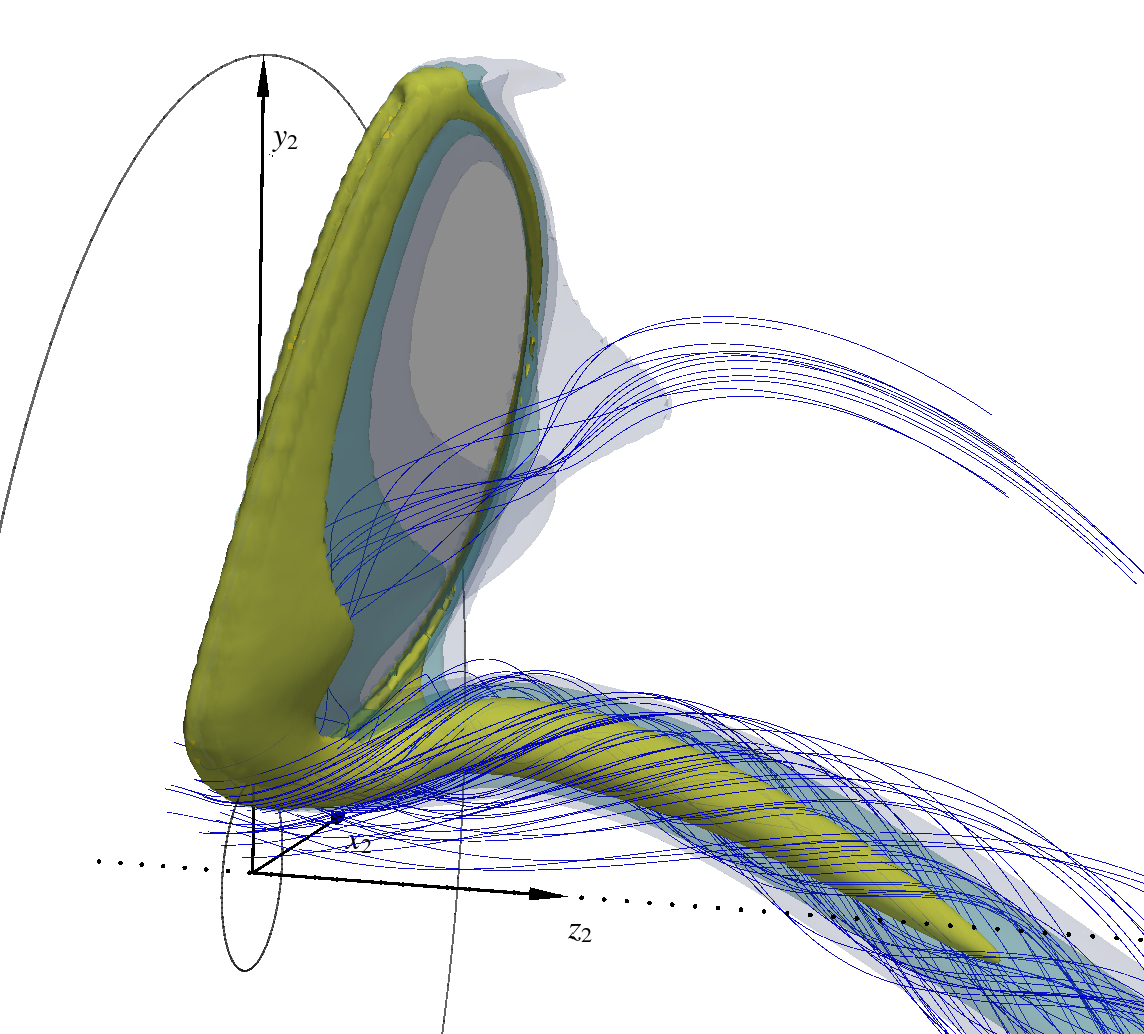}
\includegraphics[width=0.242\columnwidth, trim={100, 0, 100, 0},clip]{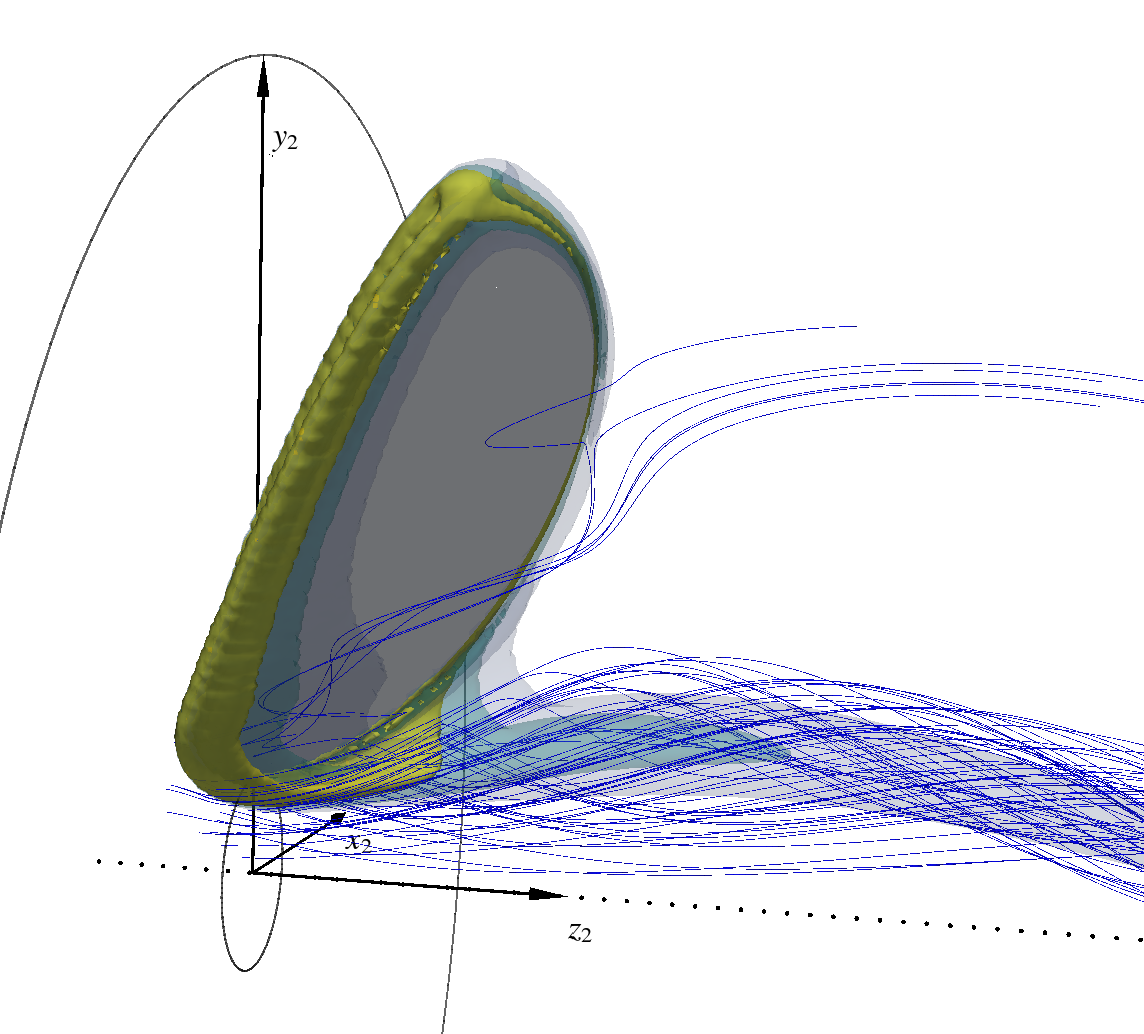}

\mylab{-0.01\textwidth}{0.29\textwidth}{$(a)$}
\mylab{ 0.23\textwidth}{0.29\textwidth}{$(b)$}
\mylab{ 0.48\textwidth}{0.29\textwidth}{$(c)$}
\mylab{ 0.72\textwidth}{0.29\textwidth}{$(d)$}
%($c$) \hspace{5cm} 
%($d$) 

\caption{
    Three-dimensional instantaneous visualizations of the relative velocity around the blade, in the non-inertial reference frame $\Sigma_2$. 
    Vortical structures are shown with coloured surfaces,  using the Q-criterion with thresholds $Qc^2/U_\infty^2 = [2, 4, 8]$.
    The relative velocity field is shown using streamlines (blue lines). 
    ($a$) Auto-rotation ($\lambda = 3.2$). 
    ($b$) Maximum power coefficient ($\lambda = 2.3$).
    ($c$) Maximum torque ($\lambda = 1.7$).
    ($d$) Lowest tip-speed ratio ($\lambda = 1$).
    \label{fig:1blade_3D}
}
%\vspace*{-0.3cm}
\end{figure}

In order to investigate the relationship between the blade's attitude and its performance in terms of $C_P$ and $M_{z,1}$, figure \ref{fig:1blade_3D} shows instantaneous flow visualizations of the flow around the blade for four characteristic braking torques: auto-rotation, maximum power, maximum torque, and minimum tip-speed ratio. 
Iso-surfaces of the second invariant of the velocity gradient tensor ($Q$) are used to visualize vortices, and the streamlines generated by a distribution of points around the leading edge (i.e, near the hinge point $J$) are used to visualize the relative velocity field.
Both velocity and $Q$ are computed in the non-inertial reference frame $\Sigma_2$, that rotates around the rotor's axis at a relatively constant $\dot\varphi$.

When the blade is in auto-rotation, figure \ref{fig:1blade_3D}a shows a strong vortex developing at the hub (yellow vortex wrapped in blue streamlines), a leading edge vortex (LEV, in yellow) covering a large fraction of the upper surface of the blade, and a shear layer emanating from the LEV (translucid green and blue isosurfaces) that rolls-up in a weak wing tip vortex at the outer edge.  
These flow structures are qualitatively similar to the flow around the winged-seed that inspired the present blade design (compare figure \ref{fig:1blade_3D}a with figure 8 in \citealp{arranz2018numerical}),  although the wing tip vortices of the present blade seem to be slightly weaker. 
Figures \ref{fig:1blade_3D}b-d show that the intensity of all vortical structures around the blade decreases when the tip-speed ratio decreases. 

\begin{figure}[t]

\begin{minipage}{0.55\textwidth}
\includegraphics[height = 0.265\textwidth,trim={0, 0, 0, 0},clip]{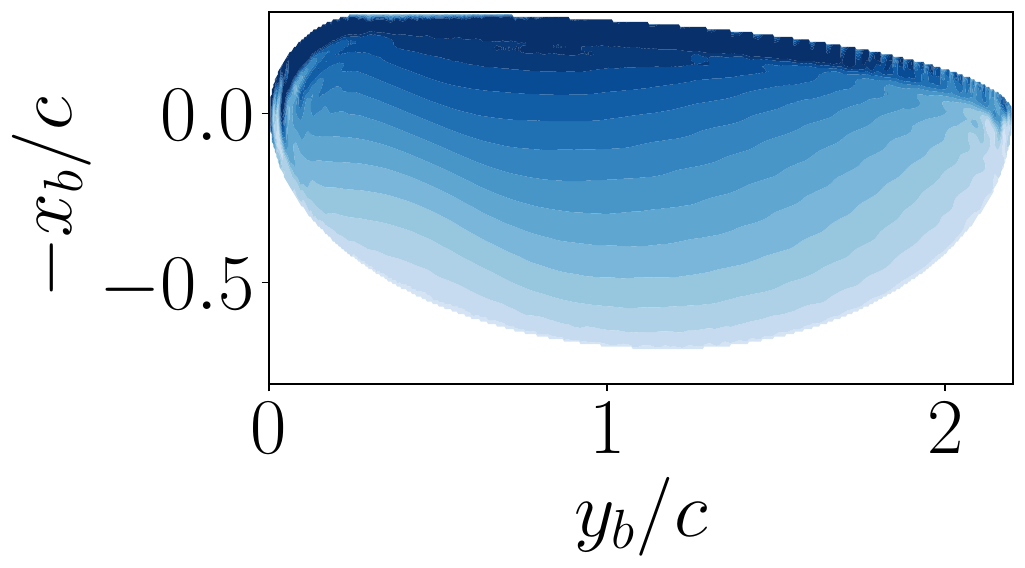}
\hspace{1mm}
\includegraphics[height = 0.265\textwidth,trim={40, 0, 0, 0},clip]{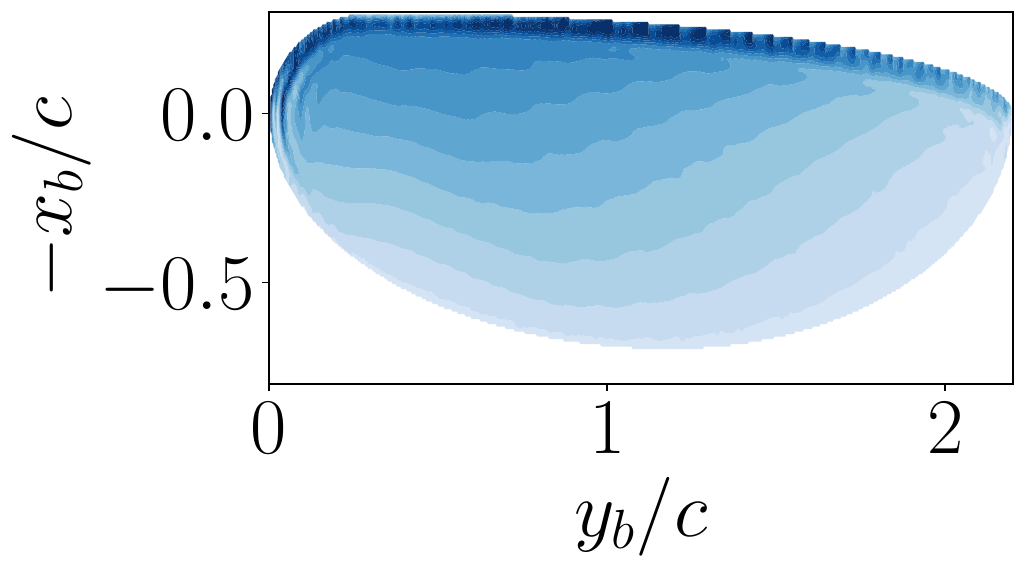}
\mylab{-0.99\textwidth}{0.24\textwidth}{($a$)}
\mylab{-0.51\textwidth}{0.24\textwidth}{($b$)}

\vspace{-3mm}
\includegraphics[height = 0.265\textwidth,trim={0, 0, 0, 0},clip]{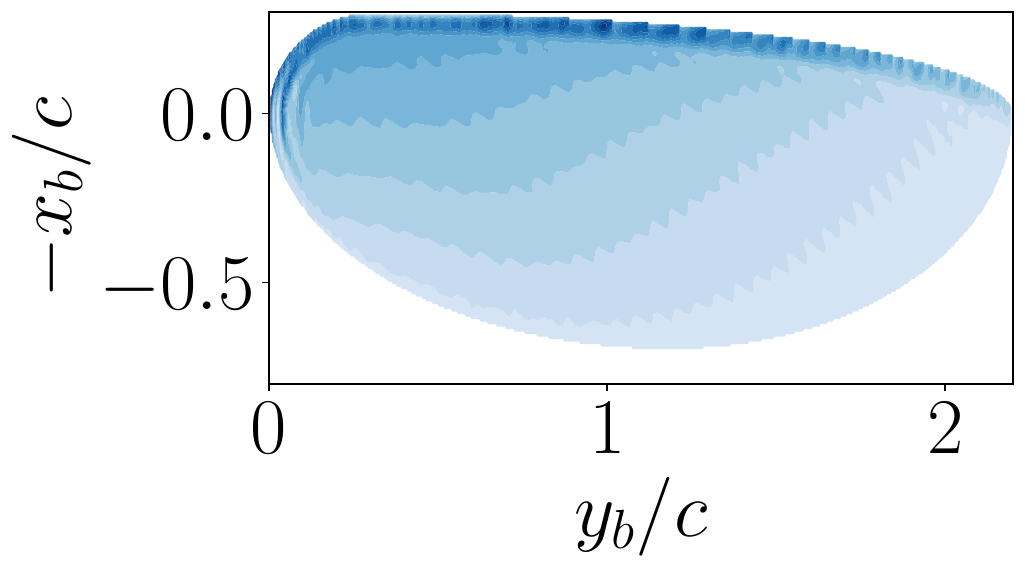}
\hspace{1mm}
\includegraphics[height = 0.265\textwidth,trim={40, 0, 0, 0},clip]{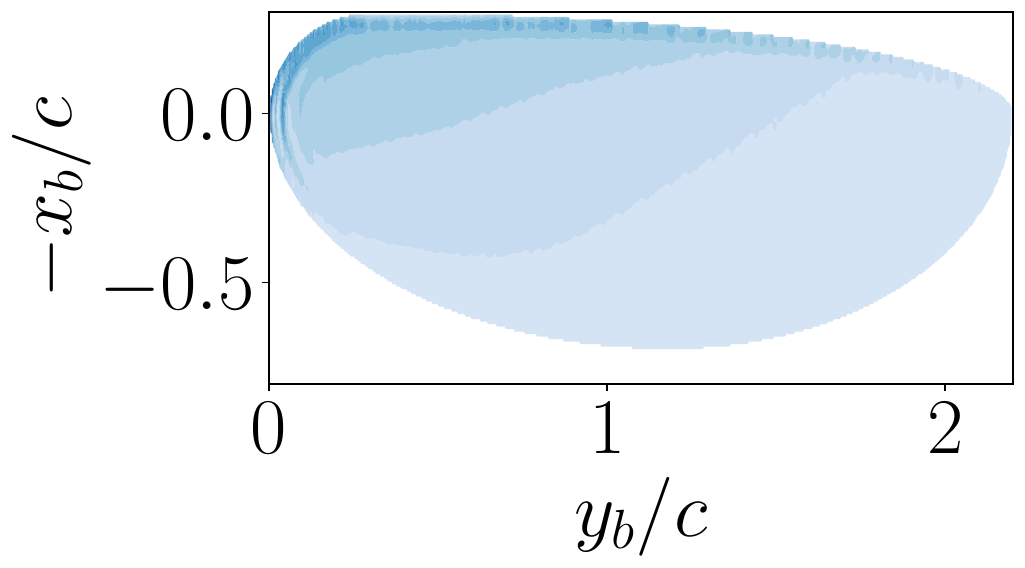}
\mylab{-0.99\textwidth}{0.24\textwidth}{($c$)}
\mylab{-0.51\textwidth}{0.24\textwidth}{($d$)}

\centering
\includegraphics[width = 0.8\textwidth]{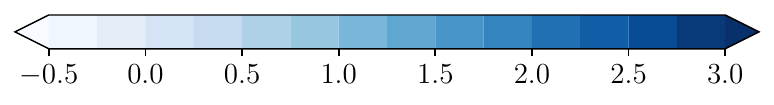}
\end{minipage}
\hspace{2mm}
\begin{minipage}{0.44\textwidth}
\includegraphics[width = 0.97\textwidth]{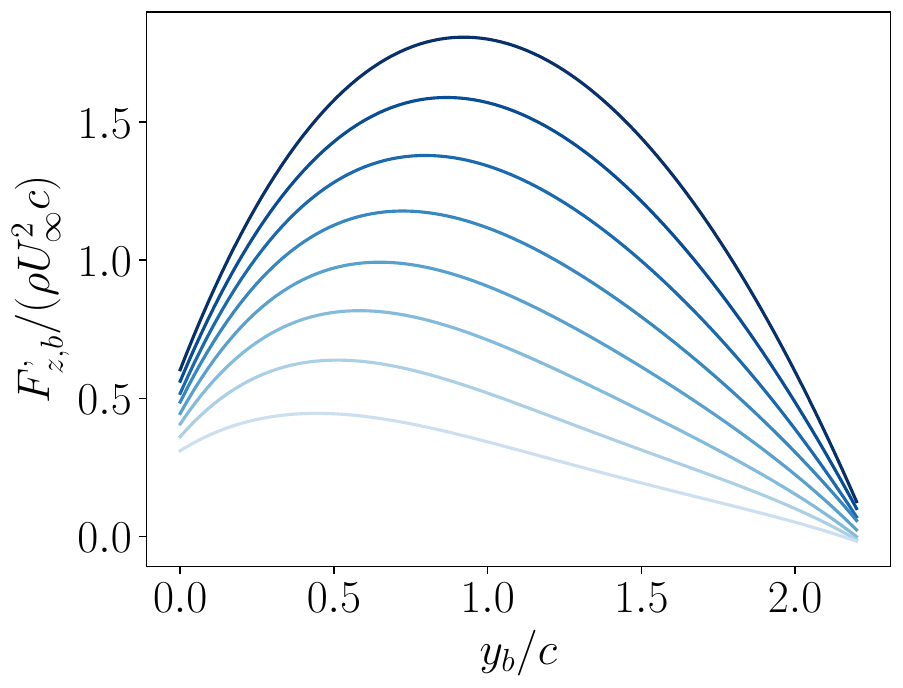}
\mylab{-1\textwidth}{0.72\textwidth}{($e$)}
\myarrow{0.6\textwidth}{0.69\textwidth}{\draw[-{Stealth[length=2.5mm,width=2mm]}] (0,0) -- (0,-2cm) node[pos=0.69]{~~~~$c_0$};}
\end{minipage}
\caption{
($a$-$d$) Distribution of the aerodynamic force normal to the blade, $f_{z,b} / (\rho U_\infty^2)$.
($a$) Auto-rotation ($\lambda = 3.2$). 
($b$) Maximum power coefficient ($\lambda = 2.3$).
($c$) Maximum torque ($\lambda = 1.7$).
($d$) Lowest tip-speed ratio ($\lambda = 1$).
($e$) Spanwise distribution of the normal force per unit span, $F'_{z,b}/ (\rho U_\infty^2 c)$. Colors as in figure \ref{fig:1blade_time}.
\label{fig:1blade_fnormal}
}
\end{figure}

The decrease on the intensity of the vortical structures shown in figure \ref{fig:1blade_3D} suggests a decrease of the pressure forces acting on the blade. 
This is evaluated in figure \ref{fig:1blade_fnormal}, which depicts the aerodynamic force normal to the surface $f_{z,b} = (x_b, y_b)$. Note that, even at the relatively low $\mathrm{Re}$ considered here, the normal component of the aerodynamic forces is approximately equal to the pressure forces, much larger than the viscous normal forces. 
Figures \ref{fig:1blade_fnormal}a-d show the distribution of $f_{z,b}$ on the blade's surface for the four representative cases considered before: auto-rotation, maximum power coefficient, maximum torque and minimum tip-speed ratio. 
Figure  \ref{fig:1blade_fnormal}e shows the spanwise distribution of the normal force per unit span, 
\begin{equation} 
F'_{z,b}(y_b) = \int_{x_{LE}}^{x_{TE}} f_{z,b}(x_b,y_b) dx_b, 
\end{equation}
where $x_{LE}(y_b)$ and $x_{TE}(y_b)$ denote the curves formed by the leading and trailing edges on the $(x_b,y_b)$ plane, respectively.   

Figures \ref{fig:1blade_fnormal}a-d show that, independently of the tip-speed ratio, the maximum normal force is always concentrated near the leading edge of the blade. 
In auto-rotation (figure~\ref{fig:1blade_fnormal}a) the sharp peak at the leading edge propagates into the blade's surface at spanwise positions 0.5 $\lesssim y_b/c \lesssim 1.5$, consistent with the shape of the LEV shown in figure \ref{fig:1blade_3D}a, and producing a spanwise distribution of normal force peaking at $y_b/c \approx 1$ (see dark blue line in figure \ref{fig:1blade_fnormal}e). 
As $c_0$ increases and the tip-speed ratio decreases, figure \ref{fig:1blade_fnormal}e shows that the intensity of the pressure forces gradually decreases, while the spanwise location of the peak in $F'_{z,b}$ moves inboard, closer to point  J.

\begin{figure}[t]

\begin{minipage}{0.55\textwidth}
\includegraphics[height = 0.25\textwidth,trim={0, 0, 0, 0},clip]{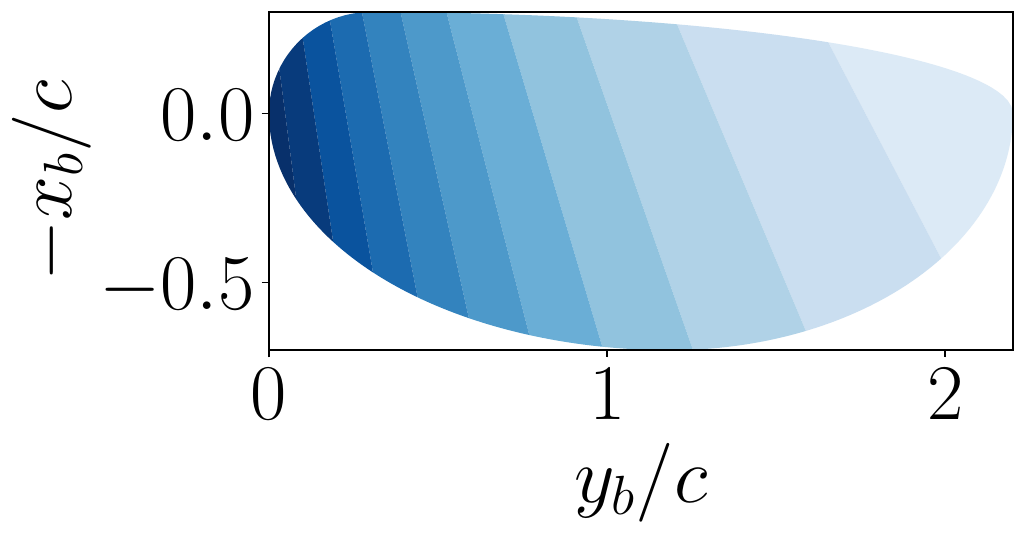}
\hspace{1mm}
\includegraphics[height = 0.25\textwidth,trim={40, 0, 0, 0},clip]{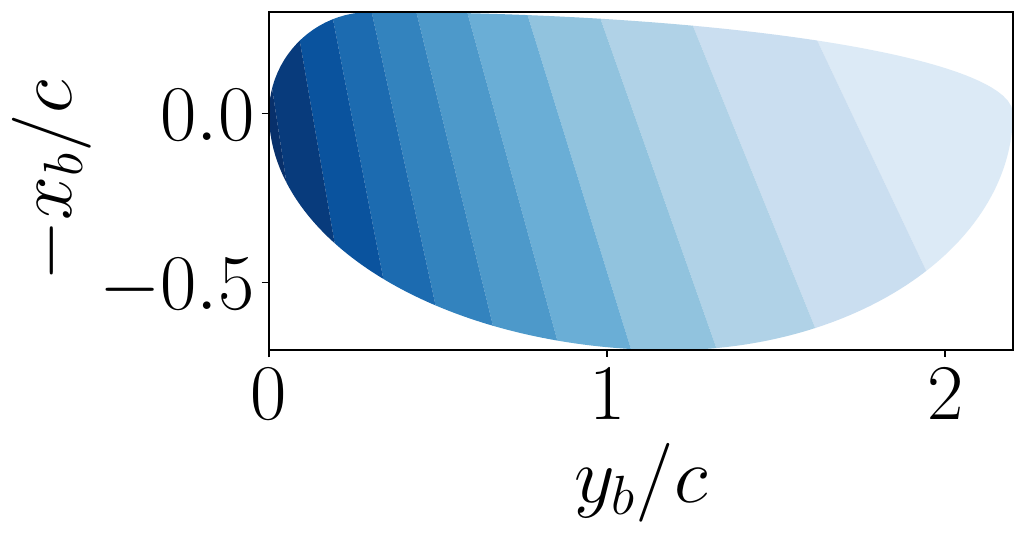}
\mylab{-0.99\textwidth}{0.24\textwidth}{($a$)}
\mylab{-0.51\textwidth}{0.24\textwidth}{($b$)}

\vspace{-3mm}
\includegraphics[height = 0.25\textwidth,trim={0, 0, 0, 0},clip]{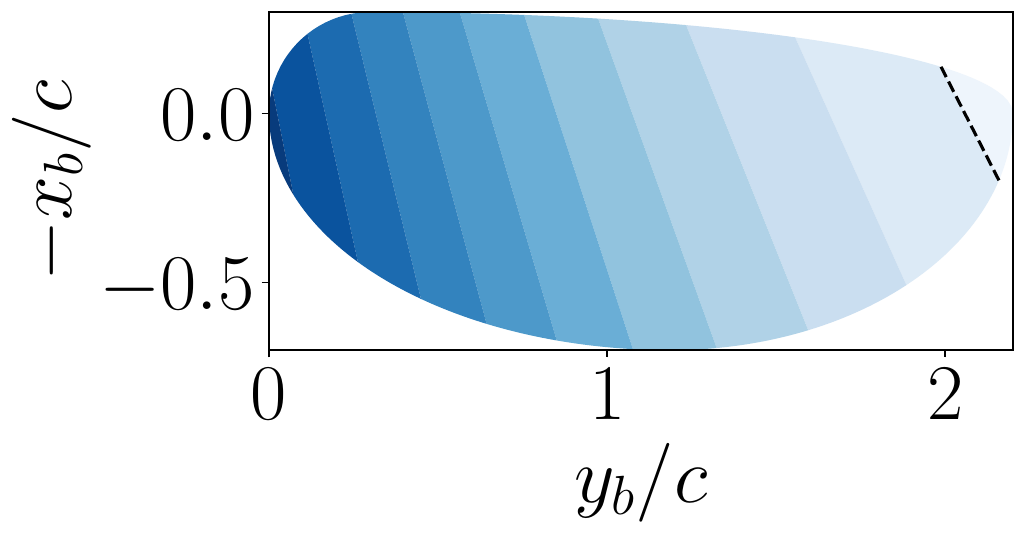}
\hspace{1mm}
\includegraphics[height = 0.25\textwidth,trim={40, 0, 0, 0},clip]{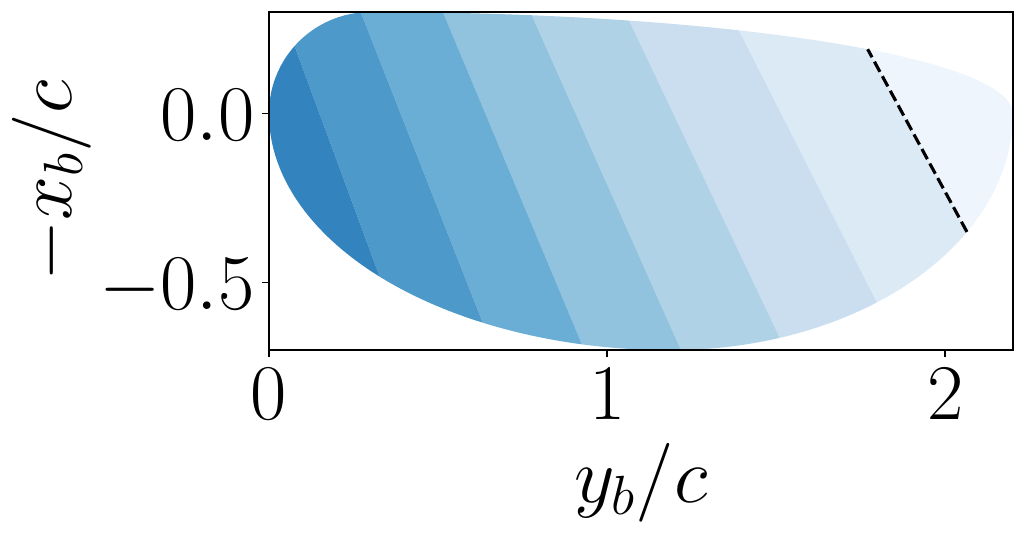}
\mylab{-0.99\textwidth}{0.24\textwidth}{($c$)}
\mylab{-0.51\textwidth}{0.24\textwidth}{($d$)}

\centering
\includegraphics[width = 0.8\textwidth]{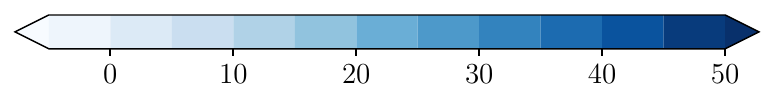}
\end{minipage}
\hspace{2mm}
\begin{minipage}{0.44\textwidth}
\includegraphics[width = 0.98\textwidth]{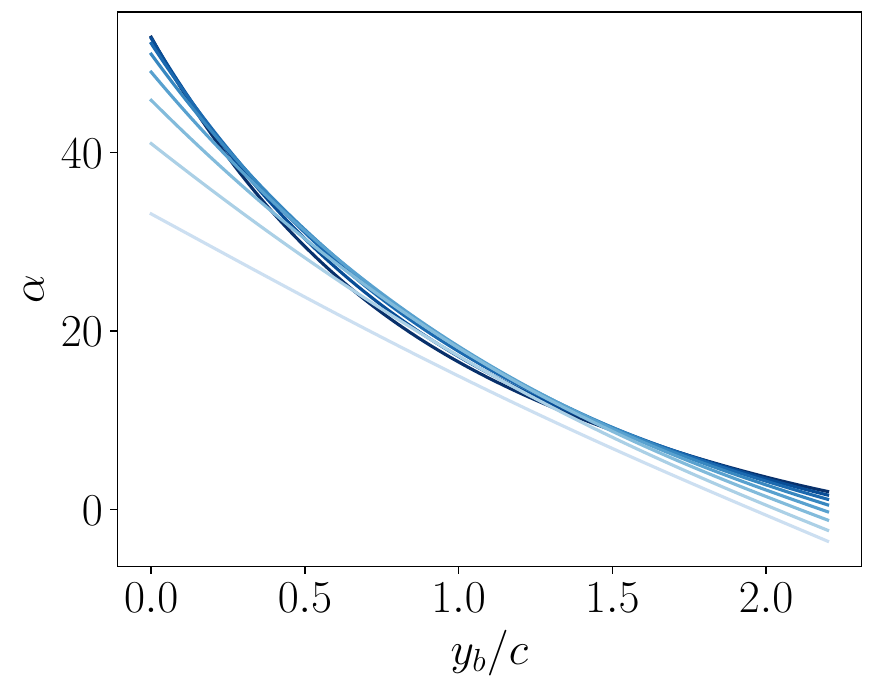}
\mylab{-1.02\textwidth}{0.73\textwidth}{($e$)}
\end{minipage}
\caption{
($a$-$d$) Distribution of the geometric angle of attack $\alpha$ on the surface of the blade. 
($a$) Auto-rotation ($\lambda = 3.2$). 
($b$) Maximum power coefficient ($\lambda = 2.3$).
($c$) Maximum torque ($\lambda = 1.7$).
($d$) Lowest tip-speed ratio ($\lambda = 1$).
Black dashed line in ($c$) and ($d$) correspond to $\alpha=0$. 
($e$) Spanwise distribution of the chordwise-averaged geometric angle of attack, $\langle \alpha \rangle$,  for all cases. 
Line colors as in figure \ref{fig:1blade_time}. 
All angles are expressed in deg. 
\label{fig:1blade_AoA}
}
\end{figure}

The gradual decrease of $F'_{z,b}(y_b)$ observed in figure \ref{fig:1blade_fnormal}e is mostly due to the decrease of dynamic pressure associated with a lower tip-speed ratio. 
Indeed, the angle of attack of the blade with respect to the incoming wind is fairly constant from auto-rotation to maximum torque, decreasing sharply for lower tip-speed ratios. 
In order to prove this statement, figure \ref{fig:1blade_AoA} presents the distributions of geometric angle of attack, $\alpha$ on the blades. 
This angle is computed as the angle between the chordwise direction of the blade (i.e., unitary vector $\mathbf{i}_b$ in the blade's reference frame $\Sigma_b$, see figure \ref{fig:samaraWT-model}a) and the $(x_b, z_b)$ components of the velocity of the blade $\mathbf U_\mathrm{blade}$ with respect to the incoming freestream. 
The latter is computed subtracting the rotational velocity of the blade from the uniform freestream.
In the steady state this yields 
\begin{equation} 
\mathbf{U}_\mathrm{blade}(x_b^P,y_b^P) = U_\infty \mathbf{k}_1 - (\dot\varphi \mathbf{k}_1) \times \overline{\mathrm{OP}}, 
\end{equation}
where $\mathbf{k}_1$ is the unitary vector along the $z_1$-direction, and $\overline{\mathrm{OP}}$ is the vector from point O to point P on the surface of the blade (with coordinates $x_b^P,y_b^P$ in the body reference frame).

As before, figures \ref{fig:1blade_AoA}a-d show the surface distribution of $\alpha$ for four representative tip-speed ratios, 
and figure \ref{fig:1blade_AoA}e shows the spanwise distribution of the chordwise-averaged value of $\alpha$ for all the braking torques considered here. 
The distributions of $\alpha$ over the blade are very similar for auto-rotation (figure \ref{fig:1blade_AoA}a), maximum power coefficient (figure \ref{fig:1blade_AoA}b) and maximum torque (figure \ref{fig:1blade_AoA}c). 
For the latter we observe however a small region of negative $\alpha$ developing near the blades outboard tip ($y_b \sim 2.5c$). 
This negative-$\alpha$ region becomes larger for case with minimum tip-speed ratio (figure \ref{fig:1blade_AoA}d), which exhibits also lower values of $\alpha$ near the hinge. 
The trend is clearer in figure \ref{fig:1blade_AoA}(e), which shows that as $\lambda$ decreases (i.e., $c_0$ increases) the changes in the blade's attitude $\beta$ and $\theta$ succeed in maintaining roughly the same distribution of $\alpha$ over most of the span, with a gradual reduction of $\alpha$ near the hinge J and near the tips. 
For tip-speed ratios below that of maximum torque the scenario changes, and $\alpha$ decreases at all spanwise sections of the blade.

It is interesting to note that the reason why $C_P$ and $M_{z,1}$ increase as $\lambda$ decreases from auto-rotation (i.e., $C_P = M_{z,1} = 0$) is the competition between two opposed mechanisms. 
On the one hand, decreasing $\lambda$ results in lower dynamic pressure and lower aerodynamic forces (figure \ref{fig:1blade_fnormal}), even if the angle of attack remains reasonable constant (figure \ref{fig:1blade_AoA}).  
On the other hand, the change in the attitude of the blade, especially the change in the pitching angle (figure \ref{fig:1blade_ST}d), results on a larger projection of the aerodynamic force on the $(x_1,y_1)$ plane, increasing the torque at the axis (even if $F_{z,b}$  is reduced). 
When $\lambda$ is too low, $\beta$ and $\theta$ render the blade's surface almost parallel to the incoming flow (figure \ref{fig:1blade_3D}d), resulting in a lower angle of attack (figure \ref{fig:1blade_AoA}d) and an even lower normal force (figure \ref{fig:1blade_fnormal}d). 
These two competing mechanisms (dynamic pressure and pressure forces decreasing with $\lambda$, and $-\theta$ increasing with $\lambda$) determine the optimal tip-speed ratio for maximum torque and maximum power, as shown in figure \ref{fig:1blade_ST}.

%
%When normalized with the freestream's dynamic pressure, the drag decreases monotonically from $F_{z,1}/(\rho U_\infty^2 c^2) = 3.0$  at $\lambda = 3.3$ (autor-rotation), to $1.2$ when the aerodynamic torque is maximum. 
%However, when normalized with the dynamic pressure of the velocity of the blade with respect to the uniform freestream (eq. \ref{eq:Ublade}) the drag coefficient remains roughly constant in that range of tip-speed-ratios. 

\subsection{Nature-inspired micro-rotor in a turbulent freestream}
\label{sec:turbulence}

%MGV: decir en algun sitio que esto es un analisis de proof of concept. 

\begin{figure}[t]
\vspace*{0.2cm}
\centering
\includegraphics[width=0.49\columnwidth]{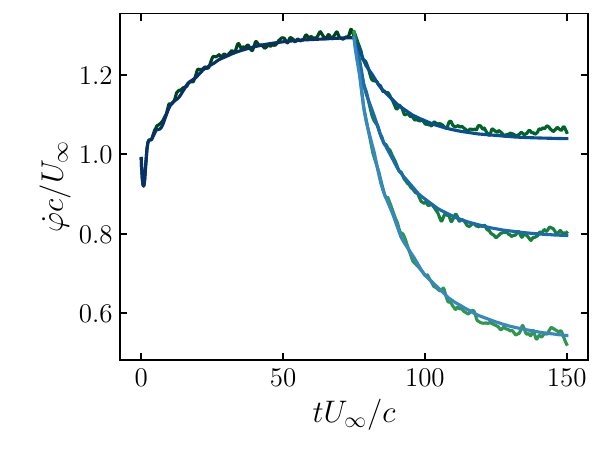}
\includegraphics[width=0.49\columnwidth]{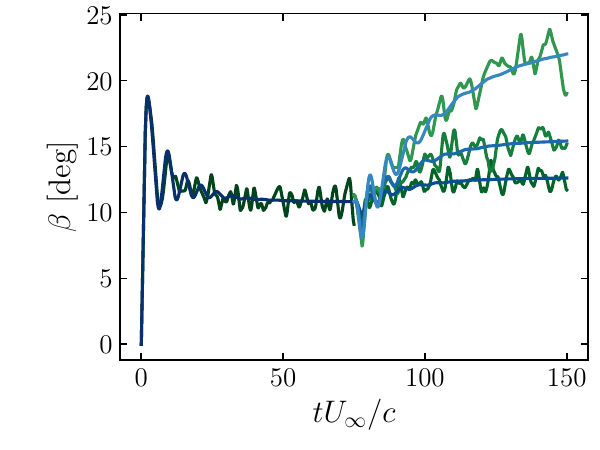}
\includegraphics[width=0.49\columnwidth]{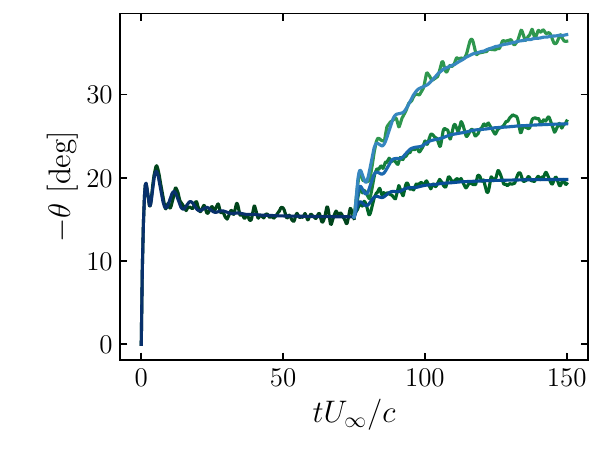}
\includegraphics[width=0.49\columnwidth]{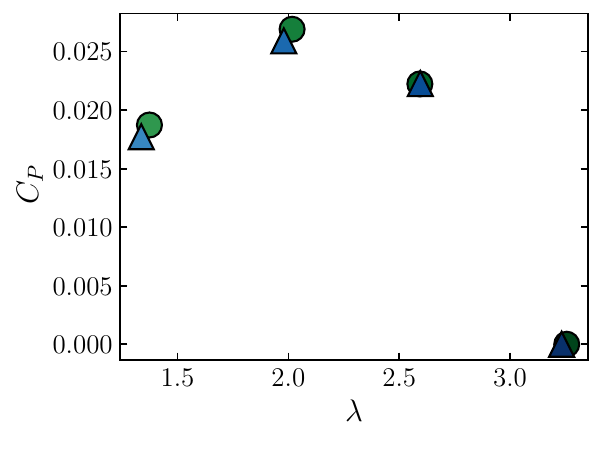}
\\
\mylab{-0.46\textwidth}{0.77\textwidth}{(a)}
\mylab{0.03\textwidth}{0.77\textwidth}{(b)}
\mylab{-0.46\textwidth}{0.4\textwidth}{(c)}
\mylab{0.03\textwidth}{0.4\textwidth}{(d)}
\caption{
Temporal evolution of the blade’s kinematics: 
(a) angular velocity,  $\dot\varphi$; (b) coning angle, $\beta$; (c) pitching angle, $\theta$. 
(d) Steady state power coefficient.
Blue/triangles for uniform free-stream, and green/circles for free-stream with turbulence-like perturbations ($\Lambda_0 = 0.5c$, $TI - 0.1$).
The intensity of the color indicate the proportionality constant of the braking torque, as in figure \ref{fig:1blade_time}. 
}
\label{fig:1blade_braking_stig}
\vspace*{-0.5cm}
\end{figure}
%------------------------------------------

Following the characterization of the single-blade nature-inspired rotor in a uniform freestream, next we turn our attention to the performance of the rotor when the incoming flow is non-uniform, as expected in operational conditions.  
To that effect, the synthetic turbulence generator described in section \ref{sec:methodology} is used to generate a free-stream of velocity $U)_\infty$ with turbulence-like fluctuations. 
These fluctuations are characterized by their turbulent intensity $TI = u'/U_\infty = 0.1$ and by their integral length scale $\Lambda_0 = 0.5c$. 
The influence region (i.e., where the $\mathbf{f}_{st}$ forcing of equation \ref{eq:fstig} is active) is located upstream of the blade, at $z_{i,st} = -3c$. 
This allows for the injected fluctuations to develop into decaying low-Reynolds turbulence before reaching the blade \citep{catalan2024generation}.

Apart from the presence of the synthetic turbulence generator, the simulations are run analogously to the simulations presented in section \ref{sec:unif-freestream}: first the rotor is run with $c_0 = 0$ until a stable auto-rotation state is reached at $t=75c/U_\infty$. This state is used as an initial condition to run the simulations of the rotor with values of the braking torque constant equal to $c_0 = 0.2, 0.4$ and $0.6$.

Figures \ref{fig:1blade_braking_stig}a-c compare the time evolution of the blade's attitude for the cases with uniform freestream (in blue) and with turbulence-like perturbations (in green). 
Overall, the agreement between both configurations is very good for the braking torques considered. 
The cases with turbulence-like perturbations show small oscillations on $\dot\varphi$ and $\theta$ (see figure \ref{fig:1blade_braking_stig}a and b, respectively), and slightly larger oscillations on the coning angle (figure \ref{fig:1blade_braking_stig}c). 
The fact that these oscillations are not enough to change the equilibrium position of the rotor suggests that the nature-inspired rotor's stability is robust. 
The power coefficient averaged over the last three cycles, $C_P$, is shown in figure  \ref{fig:1blade_braking_stig}d, using blue triangles for the uniform freestreamcase and green circles for the perturbed freestreamcase. 
Again, the agreement between both configurations is very good, further demonstrating the robustness of the performance of the nature-inspired rotor.

\begin{figure}[t]
\centering
\includegraphics[width=0.48\columnwidth]{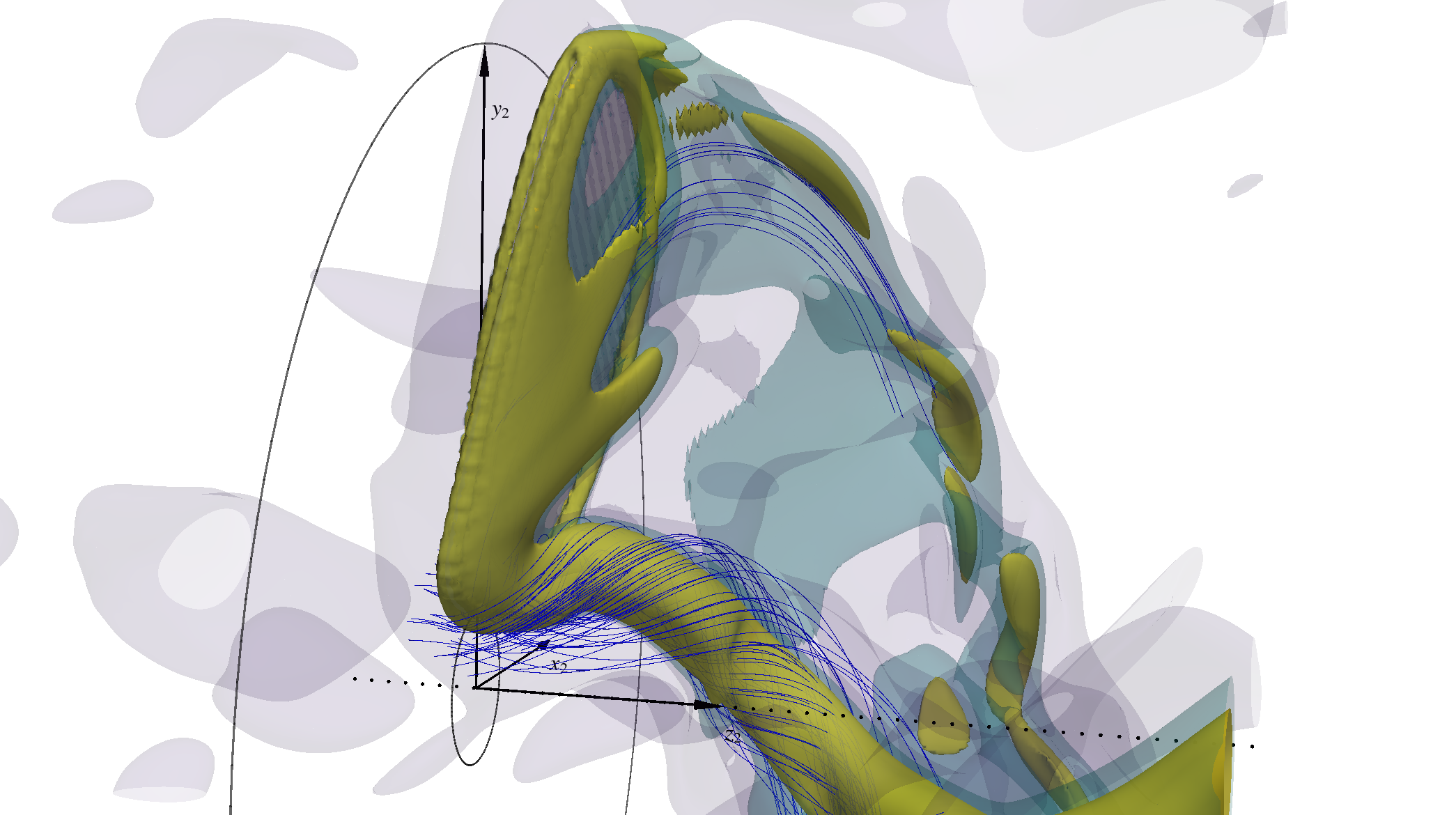}
\mylab{-0.48\textwidth}{0.26\textwidth}{($a$)}
\includegraphics[width=0.48\columnwidth]{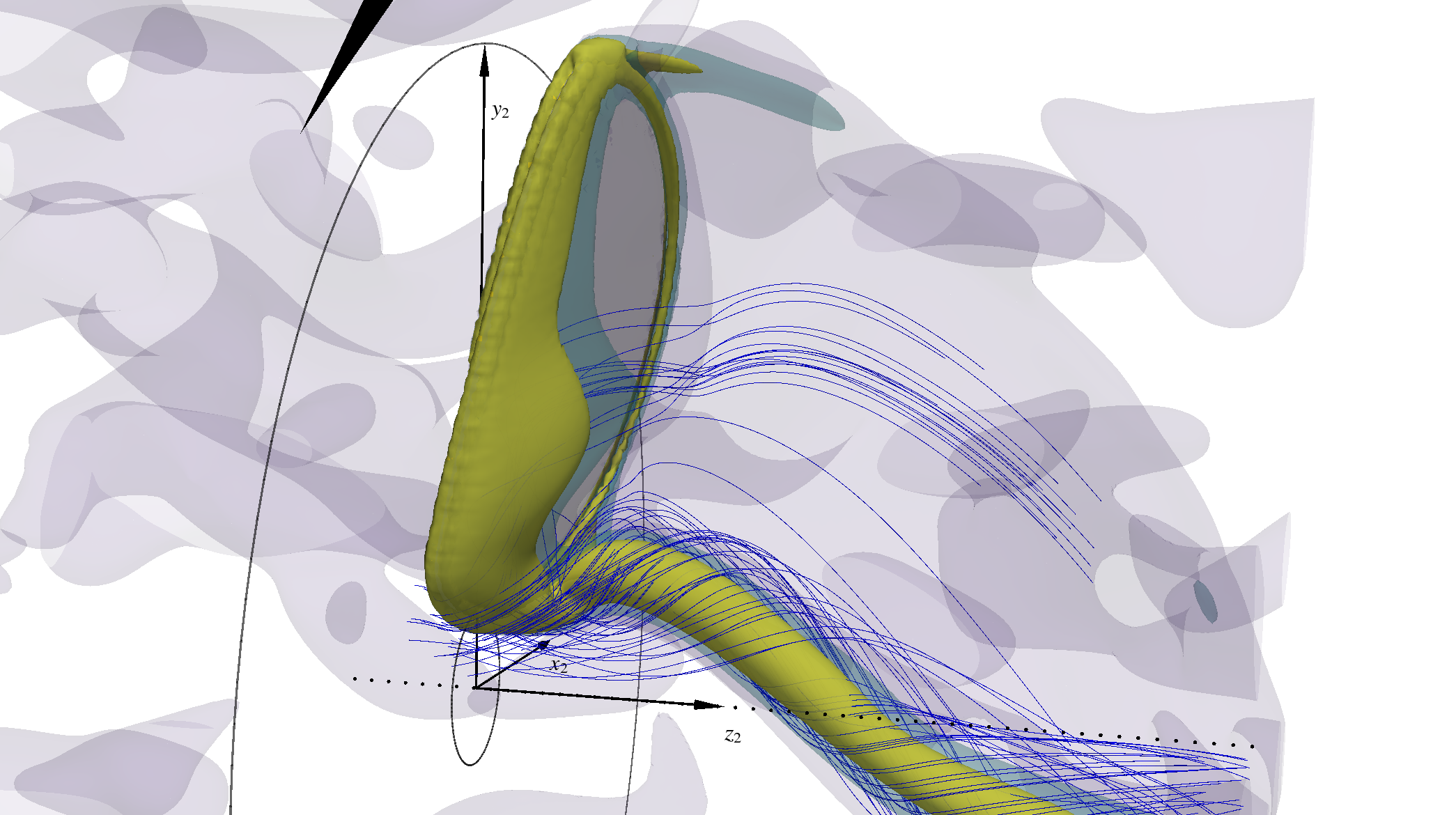}
\mylab{-0.48\textwidth}{0.26\textwidth}{($b$)}

\caption{
Three-dimensional instantaneous visualizations of the relative velocity around
the blade, in the non-inertial reference frame, $\Sigma_2$. 
Vortical structures are shown with
coloured surfaces, using the Q-criterion with thresholds, $Qc^2/U_\infty^2 = [1,4,8]$. 
The relative velocity field is shown using streamlines (blue lines).
($a$) Auto-rotation ($\lambda = 2.3$), ($b$) near-maximum power coefficient ($\lambda = 2$). 
}
\vspace*{-0.3cm}
\label{fig:stig_3D}
\end{figure}

Figure \ref{fig:stig_3D} shows the instantaneous flow features around the blade for cases with turbulent-like perturbations on the freestream.  Only two cases are shown, figure \ref{fig:stig_3D}a in auto-rotation (i.e., $c_0=0$), and figure \ref{fig:stig_3D}b around the maximum power coefficient (i.e., $c_0 = 0.4$). 
Compared to the flow structures around the blade in a uniform freestream (see figure \ref{fig:1blade_3D}), the simulations with freestream perturbations show a very similar structure of vortices, with a helicoidal vortex shed near the joint, a stable LEV on the blade's leading edge, and a weaker tip vortex. 
Note that, even if the turbulent intensity is 10\%,  the vortices developed over the blade are considerably stronger than those present in the freestream (i.e., $Q_{max} c^2/U_\infty^2\approx 1-2$ at the freestream, while the yellow iso-surface has $Q c^2/U_\infty = 8$).
This probably is the reason why the helicoidal vortex and the LEV are only slightly modified by the freestream turbulence,  
and why the performance of the rotor is not.
The results are consistent with the work of \cite{Engels2016, Engels2019} and \citet{olivieri2025}. For instance, the latter shows that a flapping wing at $\mathrm{Re}=U_\infty c/\nu = 1000$ subjected to a freestream with turbulent fluctuations virtually produces the same averaged forces as the wing in a uniform freestream.

\subsection{Nature-inspired versus fixed-blade micro-rotor}
\label{sec:bio_vs_fix}

\begin{figure}[t]
\includegraphics[width=0.45\textwidth]{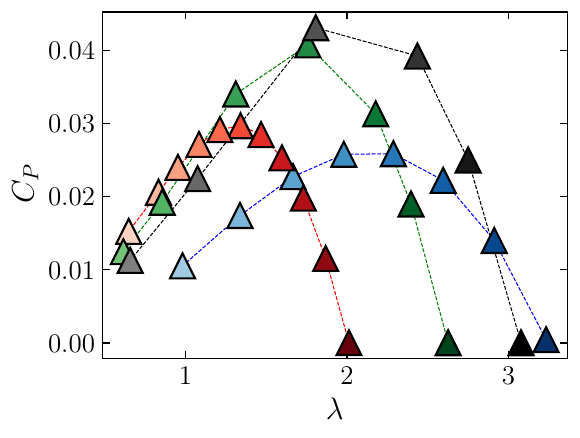}
\includegraphics[width=0.45\textwidth]{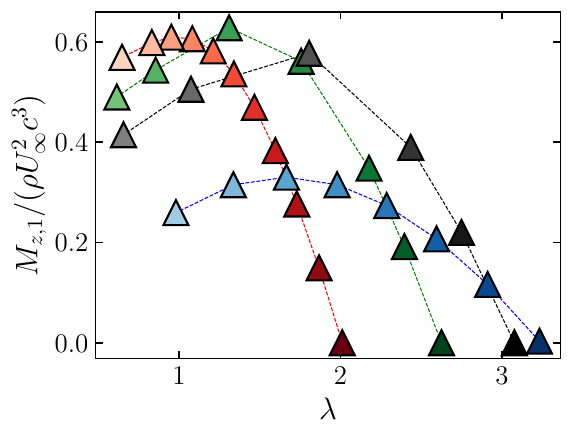}

\caption{Performance of nature-inspired rotor versus fixed angle configurations.
($a$) Power coefficient. 
($b$) Aerodynamic torque at the axis.
{\color{blue} Blue} for nature-inspired rotor; 
{\color{red} red} for FA1;
{\color{darkgreen} green} for FA2; and 
black for FA3.
\label{fig:bio-vs-fixed}
}
\end{figure}

As mentioned in section \ref{sec:unif-freestream}, the nature-inspired rotor performance is worse than the performance of other fixed-angle micro-rotors found in the literature \citep{gasnier2019, mendonca2017}. 
This comparison should be interpreted with caution, because the present simulations were conducted at a much lower Reynolds number ($\mathrm{Re} = 240$) and the current design has not been optimized. 
Recall that the blade's planform and mass distribution determine how its attitude varies with the tip-speed ratio ($\beta(\lambda)$ and $\theta(\lambda)$ in figure \ref{fig:1blade_ST}), which in turn governs the rotor's performance ($C_P(\lambda)$).

To isolate the effect of the nature-inspired versus a fixed-angle design on rotor performance, we conduct additional simulations using the same geometry and Reynolds number, but with the blade's pitching and coning angles held fixed. 
The simulations are performed as before. 
First, the blade is run in auto-rotation (i.e., $c_0=0$ and $C_P=0$) for $75c/U_\infty$ convective time units ($\approx 13$ revolutions). 
Then, a braking torque is applied (i.e., $c_0 \ne 0$), and the blade is allowed to reach a steady state. 
The torque and angular velocity are averaged over the last simulated revolution to compute $C_P(\lambda)$ for increasing values of $c_0$. 

Three fixed-blade configurations are considered and shown with different colors in figure \ref{fig:bio-vs-fixed}. 
Each configuration (FA1, FA2, and FA3) uses values of $\beta$ and $\theta$ sampled from the nature-inspired single-blade rotor discussed in section \ref{sec:unif-freestream}. 
Case FA2 (green) uses the blade attitude at $\lambda = 2.3$, corresponding to peak $C_P$. 
Cases FA1 (red) and FA3 (gray) correspond to lower and higher tip-speed ratios, $\lambda = 1.6$ and $2.9$, respectively. 
The coning and pitching angles of the fixed-angle cases are reported in table \ref{tab:fixed-angle}.

\begin{table} 
\centering
\begin{tabular}{c|c|c}
Case & $\beta$ & $\theta$ \\ 
\hline 
FA1 & $18.00^\circ$ & $-31.37^\circ$ \\
FA2 & $13.86^\circ$ & $-22.85^\circ$ \\
FA3 & $11.63^\circ$ & $-17.35^\circ$ 
\end{tabular} 
\caption{Pitching and coning angles of the fixed-angle cases.
\label{tab:fixed-angle}
}
\end{table} 

Figure \ref{fig:bio-vs-fixed}a compares power coefficient versus tip-speed ratio. 
Configurations FA2 and FA3 achieve higher $C_{P,\max}$ than the nature-inspired rotor, while FA1 reaches a similar maximum but at lower $\lambda$.  
Interestingly, the fixed-angle configurations reach their peak efficiency at smaller tip-speed ratios than the nature-inspired design, and all of them outperform it at the lowest $\lambda$. 
This suggests that the free hinge in the nature-inspired rotor allows excessive blade twisting under the incoming wind. 
Consistently, figure \ref{fig:bio-vs-fixed}b shows that the fixed-angle cases develop much larger aerodynamic torque for $\lambda \lesssim 2$ than the nature-inspired rotor, with similar peak values of $M_{z,\max} \approx 0.6 \rho U_\infty^2 c^3$ for FA1, FA2, and FA3.

\begin{figure}[t]
\includegraphics[width=0.45\textwidth]{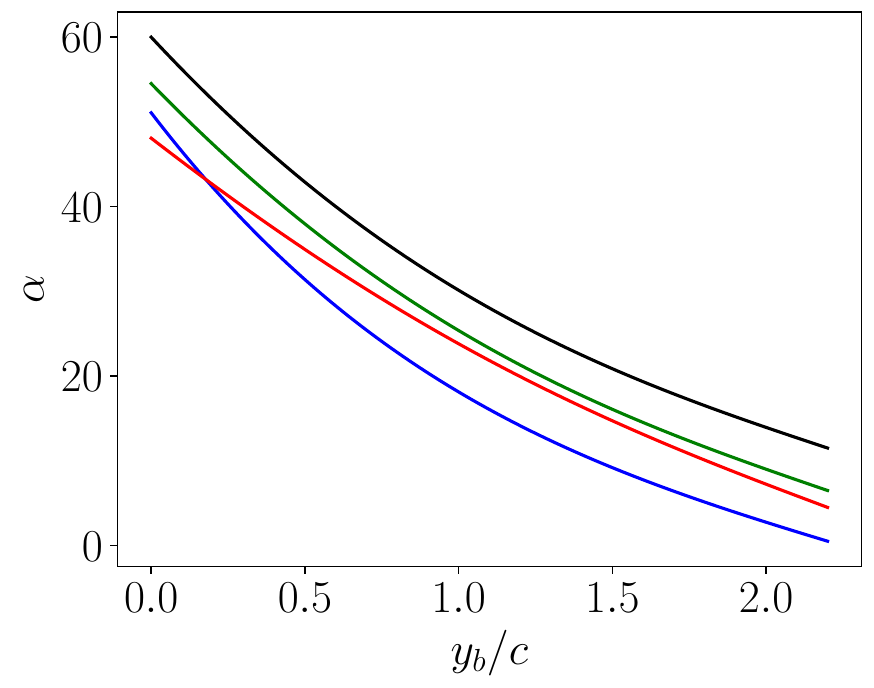}
\includegraphics[width=0.45\textwidth]{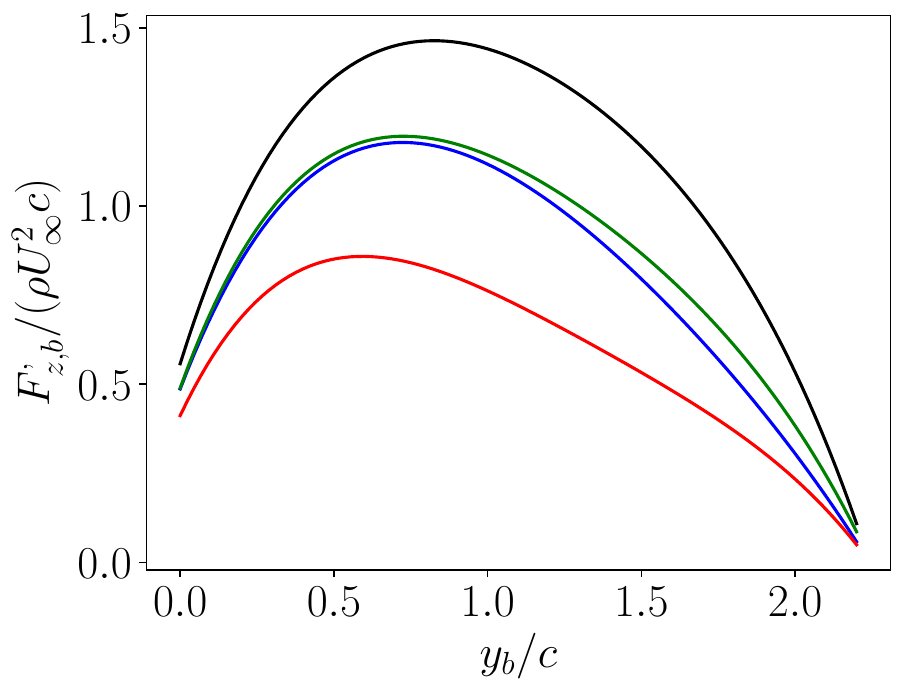}

\caption{Spanwise distributions of ($a$) angle of attack and ($b$) normal force on the blade at maximum power coefficient.  
{\color{blue} Blue} for nature-inspired rotor; 
{\color{red} red} for FA1;
{\color{darkgreen} green} for FA2; and 
black for FA3.
\label{fig:bio-vs-fixed_span}
}

\end{figure}

To explore why the fixed-angle configurations perform better, figure \ref{fig:bio-vs-fixed_span} shows spanwise distributions of angle of attack and normal force at maximum power coefficient. 
Figure \ref{fig:bio-vs-fixed_span}a indicates that fixed-angle rotors operate at larger angles of attack than the nature-inspired case. 
This generally produces stronger aerodynamic loads (figure \ref{fig:bio-vs-fixed_span}b), though the effect is modulated by changes in dynamic pressure associated with tip-speed ratio. 
Aerodynamic torque also depends on the pitching angle, which determines the projection of the normal force. 
For example, FA1 (red) has larger $\alpha$ than the nature-inspired rotor (blue) but a lower normal force due to its reduced $\lambda$, while its larger $-\theta$ increases torque. 
The role of pitching angle is even clearer when comparing FA2 (green) and FA3 (gray): although FA3 develops larger angles of attack and normal forces than FA2, its smaller pitching angle leads to a less favorable torque projection. 
As a result, FA2 and FA3 attain very similar values of $C_{P,\max}$ and $M_{z,\max}$ despite their different aerodynamic load distributions.

Overall, the comparison between the nature-inspired and fixed-angle configurations highlights the critical role of blade attitude in determining rotor performance. 
The free-hinging mechanism of the nature-inspired rotor allows for adaptive twisting, but in the current configuration it leads to suboptimal angles of attack and reduced torque at low tip-speed ratios. 
This suggests that, although the novel nature-inspired concept for a micro wind turbine explored here offers potential advantages, a proper optimization of the planform and mass distribution of the blade is necessary to match the performance of state of the art micro-rotors. 

\section{Conclusions} 
\label{sec:conclusions}

In this work we have investigated the aerodynamic performance of a novel nature-inspired micro-scale wind turbine concept, modeled after the auto-rotational flight of samara seeds. 
The micro-rotor design allows two degrees of freedom on the blade attitude: the pitch and coning (or elevation) angles. 
This enables the passive adaptation of blade attitude to the changes in the incoming flow or the rotor loading.
Direct numerical simulations of the coupled fluid–solid problem were performed at very low Reynolds number ($\mathrm{Re} = 240$), which corresponds to a micro-rotor with a chord $c=0.5$ cm, a radius $b= 1.1$cm and a free-stream velocity $U_\infty = 0.7$ m/s. 
The performance of the single-blade micro-rotor is characterized for both uniform and turbulent inflows.

Our results demonstrate that the proposed micro-rotor is able to self-regulate its attitude and sustain stable autorotation even under perturbed free-stream conditions. 
The robustness of this passive adaptation highlights the potential of our nature-inspired concept for a micro-scale wind energy harvesting, particularly in environments where ambient turbulence is unavoidable.
The leading-edge vortex and associated flow structures observed in the simulations confirm the strong similarity between the dynamics of the engineered blade and those of natural samaras.

In terms of energy harvesting performance, the present single-blade design exhibits a shallow power coefficient curve, with a maximum value of $C_{P,\max} = 0.026$ at a tip-speed ratio of $\lambda = 2.3$. 
When compared with fixed-blade configurations at the same Reynolds number, the nature-inspired rotor shows lower efficiency. 
This reduction in performance is attributed to the unoptimized blade planform and mass distribution, which cause suboptimal angles of attack at low tip-speed ratios.

Overall, the study establishes proof of concept: nature-inspired blades with free degrees of freedom can extract energy at centimeter scales and maintain stability in turbulent inflows. 
However, to realize their full potential, further optimization of the blade geometry, planform, and inertial distribution will be necessary. 
Future work will extend the present analysis to multi-blade configurations, higher Reynolds numbers, and design strategies that balance passive adaptability with improved aerodynamic efficiency.

\section*{\small Acknowledgments}
This work has benefited from fruitful discussions with Prof. C. G\"urkan about the design and performance of the nature-inspired rotor. Financial support has been provided by grant TED2021-131282B-I00 by MCIN/AEI/10.13039/501100011033 and European Union Next Generation EU/PRTR.

\appendix
%===============================================================================
\section{Grid convergence study}\label{sec:grid-convergence}

% Juanma: aqui hemos mantenido x como streamwise, no como lateral, deberiamos cambiarlo en algun momento!!!
In this section, we assess the influence of grid size by performing a grid convergence study.
To this aim, we have selected a single-blade fixed-angle rotor, with pitching and coning angles  $\theta= -16^\circ$ and $\beta=12^\circ$, at a constant angular speed $\dot{\varphi}c/U_\infty = 1.2$.  
These values are similar to those observed in our nature-inspired rotor in auto-rotation. 
The size of the computational domain ($12c\times 8c\times 8c$), the nominal Reynolds number ($\mathrm{Re}=240$) and the boundary conditions (inflow/outflow and periodic) are the same as in section \ref{sec:computational-setup}.
We have tested four values of the grid spacing $\Delta x$, corresponding to 16, 32, 48, 64 points per chord $c$.
This results in computational domains of $192\times 128 \times 128$, $384\times 256 \times 256$, $576\times 384 \times 384$ and $768 \times 512 \times 512$ grid points, respectively.
We have integrated the cases for $50c/U_\infty$ ($\lesssim 10$ cycles, where each cycle has a period of $T=2\pi/\Omega\simeq 5.23c/U_\infty$), selecting the time step $\Delta t$ accordingly to keep the CFL nearly constant between the cases.

\begin{figure}[t]
%\vspace*{0.2cm}
\centering
\includegraphics[width=0.32\columnwidth]{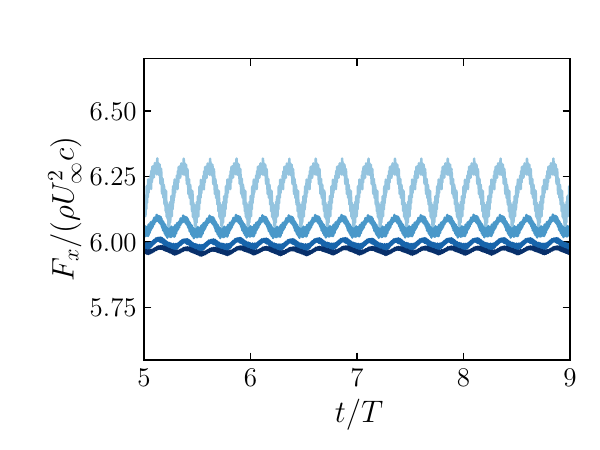}
\includegraphics[width=0.32\columnwidth]{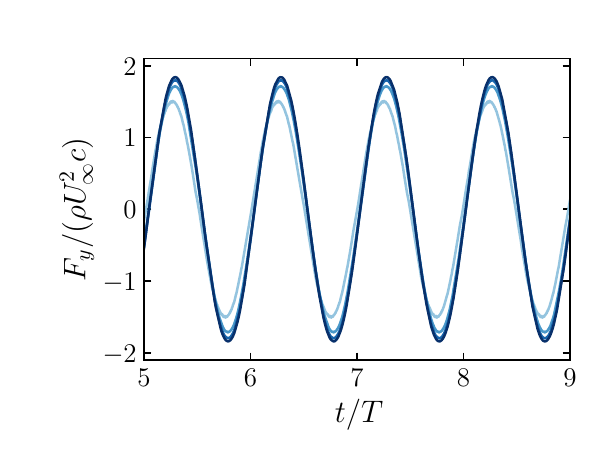}
\includegraphics[width=0.32\columnwidth]{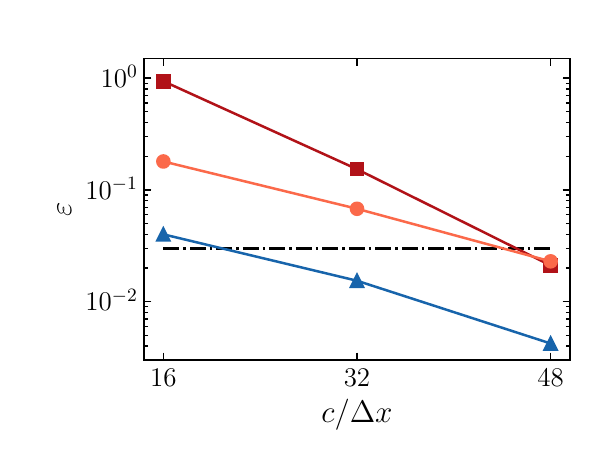}
\mylab{-0.96\textwidth}{0.22\textwidth}{(a)}
\mylab{-0.64\textwidth}{0.22\textwidth}{(b)}
\mylab{-0.31\textwidth}{0.22\textwidth}{(c)}
\caption{
    Temporal evolution of the non-dimensional axial $F_x$ (a) and lateral $F_y$ (b) forces, for several grid resolutions.
The intensity of the colors is inversely related to the grid spacing: $\Delta x=c/16$ (light blue), $\Delta x=c/32$ (sky blue), $\Delta x=c/48$ (blue), $\Delta x=c/64$ (dark blue).
Relative error $\varepsilon$ as a function of the grid resolution (c), with respect to the highest resolution ($\Delta x=c/64$).
Legend of (c): relative error in mean axial force (blue triangles), mean lateral force (light red circles) and mean peak lateral force (dark red squares).
The horizontal dashed-dotted line represents $\varepsilon = 0.03$, or $3\%$ of  error.
\vspace*{-0.5cm}
}
\label{fig:GRS}
\end{figure}

Figures \ref{fig:GRS}a,b show the averaged values of the non-dimensional axial and lateral forces, $F_x$ and $F_y$ respectively, over the last 4 cycles, while fig. \ref{fig:GRS}c displays the variation of relative error with respect to the grid spacing.
This relative error is calculated with respect to the highest resolution case ($\Delta x=c/64$). For the axial force, it is computed as:
\begin{equation}
\varepsilon_x = \frac{ |\overline{F_x}-\overline{F_x^{64}}| }{\overline{F_x^{64}} },
\end{equation}
where the $(\bar{\quad})$ operator denotes averaging over the last 4 cycles.
The same is done for the mean and mean peak lateral force.

Figures \ref{fig:GRS}a and b show that the aerodynamic forces converge as the grid spacing is decreased. 
The convergence of the axial force (figure \ref{fig:GRS}a) involves changes in the mean value and on the amplitude of the oscillations, while the convergence of the lateral force (figure \ref{fig:GRS}b) only involves changes on the amplitude of the oscillations. 
Figure \ref{fig:GRS}c shows that the resolution $c/\Delta x=32$ ensures a relative error in mean axial force around $2\%$ and a relative error in the peak lateral force of order $\mathcal{O}(10^{-1})$.
In order to bring both errors below the $3\%$ threshold, a resolution of $c/\Delta x=48$ is required. This is the grid resolution selected for the present study. 

Note that the case employed for the grid convergence analysis is the most restrictive case of our study, since the angular speed will always be lower or equal to approximately $1.2 U_{\infty}/c$.

%\bibliographystyle{plainnat}
%%\bibliographystyle{jfm}
%% Note the spaces between the initials
%\bibliography{paper}

\end{document}